\documentclass[reqno]{amsart}

\usepackage{booktabs}
\usepackage{IEEEtrantools}
\usepackage{latexsym}
\usepackage{graphicx}       
\usepackage{mathrsfs}
\usepackage{hyperref}
\usepackage{dsfont}

\usepackage[noadjust]{cite}
\usepackage{extarrows}

\usepackage{paralist}
\usepackage{capt-of}
\usepackage{xcolor}
\usepackage{diagbox}
\usepackage{cite}
\usepackage{subfigure}
\usepackage{amsmath,amssymb,amsfonts}
\usepackage{algorithmic}
\usepackage{graphicx}
\usepackage{textcomp}
\usepackage{wasysym}
\usepackage{caption}

\usepackage[ruled,vlined,linesnumbered]{algorithm2e}
\usepackage{setspace}
\usepackage{cite}
\usepackage{color}

\usepackage{color}

\definecolor{myco}{rgb}{0.55, 0.0, 0.63}

\newtheorem{theorem}{Theorem}[section]
\newtheorem{lemma}[theorem]{Lemma}
\newtheorem{problem}[theorem]{Problem}
\newtheorem{proposition}[theorem]{Proposition}
\newtheorem{corollary}[theorem]{Corollary}

\newtheorem{definition}[theorem]{Definition}
\newtheorem{example}{Example}

\newtheorem{remark}[theorem]{Remark}
\newtheorem{assumption}[theorem]{Assumption}
\numberwithin{equation}{section}   

\begin{document}
	
\begin{abstract}
In this paper, we present how to synthesize controllers to enforce $\omega$-regular properties over linear control systems affected by bounded disturbances. 
In particular, these controllers are synthesized based on so-called hybrid controlled invariant (HCI) sets.
To compute these sets, we first construct a product system between the linear control system and the deterministic Streett automata (DSA) modeling the desired property.
Then, we compute the maximal HCI set over the state set of the product system by leveraging a set-based approach.
To ensure termination of the computation of the HCI sets within a finite number of iterations, we also propose two iterative schemes to compute approximations of the maximal HCI set.
Finally, we show the effectiveness of our results via two case studies.
\end{abstract}

\title[Formal Synthesis for against $\omega$-Regular Properties: A Set-based Approach]{Formal Synthesis of Controllers for Uncertain Linear Systems against $\omega$-Regular Properties: A Set-based Approach}

\author{Bingzhuo Zhong$^{1}$}
\author{Majid Zamani$^{2,3}$}
\author{Marco Caccamo$^{1}$}
\address{$^1$TUM School of Engineering and Design, Technical University of Munich, Germany}
\email{\{bingzhuo.zhong,mcaccamo\}@tum.de}
\address{$^2$Department of Computer Science, University of Colorado Boulder, USA}
\address{$^3$Department of Computer Science, LMU Munich, Germany}
\email{majid.zamani@colorado.edu}
\maketitle

\section{Introduction}\label{sec:intro}
Formal synthesis of control systems has received significant attention in the past few years~\cite{Tabuada2009Verification} due to increasing demand for correct-by-construction controllers in many safety-critical real-life applications, such as autonomous vehicles and unmanned aerial vehicles.
These synthesis problems become more challenging when handling high-level logic properties, e.g. those expressed as linear temporal logic (LTL) formulae~\cite{Pnueli1977Temporal}, 
which are widely employed to specify properties for many applications, including~\cite{Maierhofer2022Formalization,Yu2022Distributed}.
In this paper, we focus on designing controllers enforcing $\omega$-regular properties~\cite{Thomas1990Automata}, which is a superset of LTL properties, over discrete-time linear control systems affected by bounded disturbances.

\subsection{Related Works}\label{rwork}
In the computer science community, reactive synthesis~\cite{Pnueli1989synthesis} was introduced to synthesize controllers enforcing high-level logical properties, see e.g.~\cite{Pnueli1989synthesis,Meyer2018Strix,Esparza2017LTL}.
However, these results are only applicable to systems with finite state and input sets.
As for systems with continuous state and input sets, \emph{Hamilton-Jacobi-based (HJ-based) methods}~\cite{Bansal2017Hamilton,Mitchell2005time} are applicable to synthesize controllers against invariance and reachability properties.
However, it is challenging to apply these methods to enforce high-level logic properties, in general.
To cope with high-level logic properties, 
\emph{discretization-based approaches} have been proposed in the past two decades.
Among them, \emph{symbolic techniques} (see e.g. \cite{Zamani2011Symbolic,Pola2008Approximately,Reissig2017Feedback}) are widely applied for various types of properties, such as (safe-)LTL (see e.g.~\cite{Rungger2013Specification,Tabuada2006Linear}) and $\omega$-regular properties (see e.g.~\cite{Dutreix2020Abstraction,Khaled2021OmegaThreads}).
These techniques require the construction of symbolic models (a.k.a. finite abstractions) with finite state and input sets for the original systems.
Since the  finite state and input sets are constructed by gridding the original sets, the number of discrete states and inputs grow exponentially with respect to the dimensions of state and input sets, respectively.
This issue is known as the~\emph{curse of dimensionality}, which is one of the main challenges of discretization-based approaches.
Some recent results alleviate this issue partially by constructing abstractions in a compositional manner (see e.g.~\cite{Swikir2019Compositional,Zamani2018Compositional,Pola2017Decentralized}), by leveraging a counterexample-guided abstraction refinement framework~\cite{Wolff2013Automaton}, or by applying a specification-guided framework (see e.g.~\cite{Zibaeenejad2019Auditor,Dutreix2020Specification,Meyer2018Compositional}).
However, these results require either specific properties of the systems (e.g. dissipativity, mixed-monotonicity, etc.), or additional assumptions regarding the properties (e.g. properties can be decomposed into several simpler ones).

Recently, other discretization-based approaches, which are developed based on interval analysis (referred to as \emph{interval-analysis-based approaches}), have been proposed to enforce invariance properties~\cite{Li2017Invariance}, reach-and-stay properties~\cite{Li2020Robustly}, and properties modeled by deterministic B\"{u}chi automaton~\cite{Li2022Specification}, which are subsets of $\omega$-regular properties. 
Despite improvements in terms of space complexity compared with the symbolic techniques, interval-analysis-based approaches also suffer from the curse of dimensionality, since discretization of the state sets is still needed.
Additionally, they are only applicable to systems without exogenous disturbances.

To avoid the curse of dimensionality introduced by discretizing the state and input sets, some~\emph{discretization-free approaches} have been proposed. 
Results in~\cite{Bertsekas1972Infinite} propose a set-based approach to enforce invariance properties (i.e. the systems are expected to stay within a set).
This result is further extended in~\cite{Rakovic2004Computation,Blanchini2015Set,Rungger2017Computing,Liu2019Compositional} in terms of termination and compositionality.
\emph{Control barrier functions} (CBF)~\cite{Wieland2007Constructive} are also used to enforce invariance properties (e.g.~\cite{Jahanshahi2021Compositional,Nejati2020Compositional,Jankovic2018Control,Ames2016Control}), properties described by deterministic finite automata~\cite{Jagtap2020Formal,Anand2021small}, deterministic B\"{u}chi automata~\cite{Jagtap2020Compositional}, LTL~\cite{Srinivasan2018Control}, and $\omega$-regular properties~\cite{Anand2021Compositional}.
Unfortunately, constructing valid CBFs is an NP-hard problem in general~\cite{Choi2021Robust}.

\subsection{Contribution}
In this paper, we propose new discretization-free approaches for synthesizing controllers against $\omega$-regular properties over discrete-time linear control systems affected by bounded disturbances.
Concretely, we develop set-based approaches that leverage iterative schemes to compute so-called hybrid controlled invariant (HCI) sets.
Based on these sets, one can construct controllers enforcing the desired $\omega$-regular properties.
A limited subset of the results in this paper has been presented in~\cite{Zhong2022Compositional}.
In this work, we provide detailed proofs for the results in~\cite{Zhong2022Compositional}, which are omitted in~\cite{Zhong2022Compositional}.
Additionally, we generalize the results in~\cite{Zhong2022Compositional} for synthesizing the HCI-based controllers (c.f. Remark~\ref{re:improve}), and consider an additional iterative scheme for computing under-approximations of the maximal HCI sets.
If the maximal HCI set exists, we show that these approximations can be obtained within a finite number of iterations using both iterative schemes.
Moreover, here, we show that one can obtain approximations that are arbitrarily close to the maximal HCI sets.
Finally, we provide a worst-case complexity analysis for the proposed set-based approaches.

Here, we also compare our approaches with existing results in the literature (see detailed discussion above in the related works).
In comparison with those discretization-based approaches, we do not need to discretize the state and input sets, so that our proposed approaches can be efficient in some cases in terms of computation time (c.f. Section~\ref{sec:compare}).
Compared with the discretization-free approaches based on CBFs, our proposed methods are more systematic in the sense that given any linear control systems and desired $\omega$-regular properties, one can readily compute the HCI sets and construct their corresponding controllers by leveraging our results.
Meanwhile, the results in~\cite{Anand2021Compositional} tackle the synthesis problems by first decomposing the synthesis task for the original property into several simpler ones and then computing CBFs for each simpler task by solving a series of sum-of-square (SOS) optimization problems. 
For each SOS optimization problem, one needs to choose the forms of the potential CBFs to be polynomials of fixed degrees and fix the forms of their corresponding controllers heuristically, which requires much manual effort.

\subsection{Organization}
The remainder of this paper is structured as follows.
In Section~\ref{sec2}, we provide preliminary discussions on notations, models, and the underlying problems to be solved.
Then, we discuss in Section~\ref{sec3} how to solve the synthesis problem by leveraging a set-based approach for computing HCI sets.
We provide in Section~\ref{sec4} two iterative approaches to approximate the maximal HCI sets within a finite number of iterations. 
Afterwards, we analyze the complexity of our approaches in Section~\ref{sec:compl}.
Finally, we apply our methods to two case studies in Section~\ref{sec5} and conclude our work in Section~\ref{sec6}.

\section{Notations and Preliminaries}\label{sec2}
\subsection{Notations}
We use $\mathbb{R}$ and $\mathbb{N}$ to denote the sets of real and natural numbers, respectively. 
These symbols are annotated with subscripts to restrict the sets in a usual way, e.g. $\mathbb{R}_{\geq0}$ denotes the set of non-negative real numbers.
Moreover, $\mathbb{R}^{n\times m}$ with $n,m\in \mathbb{N}_{\geq 1}$ denotes the vector space of real matrices with $n$ rows and $m$ columns.
For $a,b\in\mathbb{R}$ (resp. $a,b\in\mathbb{N}$) with $a\leq b$, the closed, open, and half-open intervals in $\mathbb{R}$ (resp. $\mathbb{N}$) are denoted by $[a,b]$, $(a,b)$ ,$[a,b)$, and $(a,b]$, respectively. 
We denote by $\mathbf{0}_n$ and $\mathbf{I}_n$ the column vector in $\mathbb{R}^n$ with all elements equal to 0, and the identity matrix in $\mathbb{R}^{n\times n}$, respectively. 
Given $N$ vectors $x_i \in \mathbb R^{n_i}$, $n_i\in \mathbb N_{\ge 1}$, and $i\in\{1,\ldots,N\}$, we use $x = [x_1;\ldots;x_N]$ to denote the corresponding column vector of dimension $\sum_i n_i$.
Additionally, given a vector $x\in\mathbb{R}^n$, we denote by $|x|$ and $\lVert x \rVert$ the infinity and Euclidean norm of $x$, respectively.
We denote by $\mathbb{B}^n$ the closed unit ball centered at the origin in $\mathbb{R}^n$ with respect to the infinity norm.
Given sets $A$ and $B$, we denote by $f:A \rightarrow B$ an ordinary map from $A$ to $B$.
Given sets $X_i$, $i\in[1,N]$, and their Cartesian product $X_1\times \ldots \times X_N$, 
the projection of $X$ onto $X_i$ is denoted by mapping $\pi_{X_i}: X\rightarrow X_i$.
Considering a set $\Pi$, $\Pi^{\omega}$ denotes the Cartesian product of an infinite number of $\Pi$.
Given sets $A$ and $B$ with $A\subset B$, $B\backslash A = \{x|x\in B\ \text{and}\ x\notin A\}$ denotes the complement of $A$ with respect to $B$.
The Minkowski sum of two sets $A$, $B\subseteq \mathbb{R}^n$ is denoted by $A+B=\{x\in \mathbb{R}^n| \exists a\in A,\,\exists b\in B,\, x=a+b\}$.
In this paper, we slightly abuse the notation and use $x+A$ instead of $\{x\}+A$ to denote the Minkowski sum of set $A$ and $\{x\}$ where $x\in\mathbb{R}^n$.
Moreover,  $A-B=\{a\in A|a+B\subseteq A\}$ denotes the Pontryagin set difference between $A$ and $B$. 

\subsection{Systems}
In this paper, we focus on discrete-time linear control systems (dtLCS), which are defined as follows.
\begin{definition}\label{def:dtLCS}
	(\emph{dtLCS}) A discrete-time linear control system $S$ is a tuple
	\begin{align}
	S = (X,X_0,U,W,f),\label{sys:tuple}
	\end{align}
	where $X\!\subseteq\! \mathbb{R}^{n}$ is the state set, $U\!\subset\! \mathbb{R}^{m}$ and $W\!\subset\!\mathbb{R}^{n}$ are compact sets of input and exogenous disturbances, respectively.
	Set $X_0\subseteq X$ is the set of initial states.
	Function $f:\! X\!\times\! U \!\times\! W\!\rightarrow\! X$ characterizes the discrete-time dynamics  as: 
	\begin{equation}
	x(k+1) = f(x(k),u(k),w(k)):=Ax(k)+Bu(k)+w(k),\label{eq:linear_subsys}
	\end{equation}
	with $A\in \mathbb{R}^{n\times n}$ and $B \in \mathbb{R}^{n\times m}$.
\end{definition}
With these notations, the evolution of the system $S$ as in~\eqref{sys:tuple} can be described by its paths, as defined below.
\begin{definition}
	(\emph{Path})
	A path of a dtLCS $S$ as in~\eqref{eq:linear_subsys} is 
	\begin{align*}
	\xi\,:=\,(x(0),u(0),\ldots,x(k-1),u(k-1),x(k),\ldots),k\in\mathbb{N}
	\end{align*}
	where $x(k+1)=Ax(k)+Bu(k)+w(k)$ for some $w(k)\in W$.
\end{definition}
Moreover, we denote by $\xi_{x}:=(x(0),x(1),\ldots,$ $x(k),\ldots)$ and $\xi_{u}:=(u(0),u(1),\ldots,u(k),\ldots)$ the subsequences of states and inputs in $\xi$, respectively.
Next, we proceed with defining the properties of interest.

\subsection{$\omega$-Regular Properties}\label{sec:omega_property}
The main goal of this work is to synthesize controllers enforcing $\omega$-regular properties over discrete-time linear control systems.
These properties can be modeled by deterministic Streett automata (DSA)~\cite{Streett1982Propositional}, as defined below.

\begin{definition}\label{def:DSA}
	A DSA is a tuple $\mathcal{A} = (Q, q_0, \Pi, \delta, \text{Acc})$, where $Q$ is a finite set of states, $q_0\in Q$ is an initial state, $\Pi$ is a finite set of alphabet, $\delta \subseteq Q \times \Pi \times Q$ is the set of all feasible transitions among $Q$, and $\text{Acc}=\{\langle E_1,F_1\rangle,\langle E_2,F_2\rangle,\ldots,\langle E_r,F_r\rangle,\ldots,\langle E_{\mathsf{r}},F_{\mathsf{r}}\rangle\}$ denotes the accepting condition of the DSA where $\langle E_r,F_r\rangle$, $\forall r \in\{1,\ldots,\mathsf{r}\}$, are accepting state set pairs, with $E_r,F_r\subseteq Q$.
\end{definition}
Consider an infinite word denoted by $\sigma = (\sigma_0, \sigma_1,\ldots)\in \Pi^{\omega}$.
An infinite \emph{state run} $\mathbf{q}=(q_0,q_1,\ldots)\in Q^{\omega}$ on $\sigma$ is an infinite sequence of states in which one has $(q_k,\sigma_k,q_{k+1})\in \delta$, $\forall k \in\mathbb{N}$.
Similarly, consider an finite word denoted by $\sigma_f = (\sigma_0, \ldots,\sigma_{H})\in \Pi^{H+1}$, with $H\in \mathbb{N}$, we denote by $\mathbf{q}=(q_0,\ldots,q_{H})\in Q^{H+1}$ the corresponding finite state run.
An infinite \emph{run} $\mathbf{q}$ is an \emph{accepting run} of $\mathcal{A}$, if for all $\langle E_r,F_r\rangle\,\in\text{Acc}$, $r \in\{1,\ldots,\mathsf{r}\}$, one has 
\begin{align}
inf(\mathbf{q})\,\wedge\,E_r=\emptyset \text{ or } inf(\mathbf{q})\,\wedge\,F_r \neq \emptyset,\label{acc_cond}
\end{align}
where 
$inf(\mathbf{q})$ is the set of states in $Q$ that are visited infinitely often in $\mathbf{q}$.
Additionally, an infinite word $\sigma$ corresponding to an accepting run $\mathbf{q}$ is said to be \emph{accepted by} $\mathcal{A}$.
The set of words accepted by $\mathcal{A}$, denoted by $\mathcal{L}(\mathcal{A})$,  is called the \emph{language of $\mathcal{A}$}.
Next, we define a \emph{labeling function}, which is used to connect a system $S$ as in~\eqref{sys:tuple} to a DSA $\mathcal{A}$.

\begin{definition}\label{def:L-function}
	\emph{(Labeling function)}
	Consider a dtLCS $S=(X,X_0,U,W,f)$ and a DSA $\mathcal{A} = (Q, q_0, \Pi, \delta, \text{Acc})$.
	We define a measurable labeling function $L: X\rightarrow \Pi$ as follows:
	given an infinite state sequence $\xi_{x}=(x(0),x(1),\ldots)\in X^{\omega}$ of system $S$, the word of $\xi_{x}$ over $\Pi$ is $L(\xi_{x}) = (\sigma_0,\sigma_1,\ldots,\sigma_k,\ldots)$, where $\sigma_k=L(x(k))$ for all $k\in\mathbb{N}$.
	Accordingly, we denote by $L(\xi_{x})\models \mathcal{A}$ if $L(\xi_{x})\in \mathcal{L}(\mathcal{A})$, and by $S\models\mathcal{A}$, if $L(\xi_{x})\models \mathcal{A}$ holds for all possible $\xi_{x}$ of $S$.
\end{definition}

Note that in Definition~\ref{def:L-function}, we slightly abuse the notation by applying the map $L(\cdot)$ over the domain $X^{\omega}$, i.e. $L((x(0),x(1),\ldots)) = (L(x(0)),L(x(1)),\dots)$.
However, the distinction is clear from the context.
It is also worth noting that the set of alphabet $\Pi = \{\sigma_1,\sigma_2,\ldots,\sigma_M \}$ along with the labeling function $L:X\rightarrow \Pi$ provide a partition of the state set $X=\cup^{M}_{j=1}X_j$, where $X_j:=L^{-1}(\sigma_j)$.
Finally, we propose two additional definitions related to the \emph{strongly connected components}~\cite{Baier2008Principles} in a DSA.
	\begin{definition}
		Consider a DSA $\mathcal{A} \!=\! (Q, q_0, \Pi, \delta, \text{Acc})$.
		A set $Q_1\!\subseteq \!Q$ is \emph{strongly connected} if any arbitrary pair of states $q_a,q_b\!\in\! Q_1$ are mutually reachable, i.e. $\exists (q_a,\ldots,q_b)\!\in\! Q^{d_1}, (q_b,\ldots,q_a)\in Q^{d_2}$ with $ d_1, d_2 \!\in\! \mathbb{N}$.
		A set $Q_1 \subseteq Q$ is a \emph{strongly connected component} in $\mathcal{A}$ if $Q_1$ is strongly connected, and $\nexists Q_2\!\subseteq\! Q$, with $Q_1\!\subset\! Q_2$, such that $Q_2$ is strongly connected.
		Additionally, we denote by $\text{SCC}(\mathcal{A})\subset 2^Q$ the set of all strongly connected components in $\mathcal{A}$.
	\end{definition}
	\begin{definition}\label{def:rdDRA}
		(\emph{reduced DSA})
		Consider a DSA $\mathcal{A} = (Q, q_0, \Pi, \delta, \text{Acc})$.
		A \emph{reduced DSA} of $\mathcal{A}$ with respect to a set $\bar{Q}\subset Q$  is defined as $\mathcal{A}_{rd}(\bar{Q}):=(Q', q_0, \Pi', \delta', \text{Acc}')$, with 
		$Q'\subseteq Q$, $\Pi'\subseteq \Pi$, $\delta'\subseteq \delta$, and $\text{Acc}'\subseteq \text{Acc}$ such that $\forall Q_{scc}\in \text{SCC}(\mathcal{A}_{rd}(\bar{Q}))$, $\nexists q\in\bar{Q}$ such that $q\in Q_{scc}$.
\end{definition}
Intuitively, the reduced DSA $\mathcal{A}_{rd}(\bar{Q})$ is constructed such that it does not have any strongly connected component containing the state within the set $\bar{Q}$.
So far, we have formally defined desired properties.
Next, we formulate the main problem we plan to solve in this work.

\subsection{Problem Formulation}
To formulate the main problem, we need the following definitions, which are borrowed from~\cite{Gruenbaum2003Convex}.
\begin{definition}
	(\emph{Hyperplane}) A hyperplane in $\mathbb{R}^n$ is a set
	\begin{equation}
	\{x\in\mathbb{R}^n| a^Tx=b \},
	\end{equation}
	where $a\in\mathbb{R}^n$ is non-zero and $b\in\mathbb{R}$.
\end{definition}
\begin{definition}\label{def:poly}
	A \emph{Polytope} is a bounded set of the form
	\begin{equation}
	\mathcal{P}= \{x \in \mathbb{R}^n|Px\leq p \},\label{eq:polytope}
	\end{equation}
	with $P\!\in\!\mathbb{R}^{n_p \times n}$, $p\!\in\!\mathbb{R}^{n_p}$, and $n_p\!\in\!\mathbb{N}$, where the inequality in~\eqref{eq:polytope} is component-wise.
	Accordingly, we denote by
	\begin{equation}
	\mathsf{numh}(\mathcal{P}):=n_p,\label{eq:numh}
	\end{equation}
	the number of hyperplanes defining $\mathcal{P}$, and denoted by $\mathcal{P}(n)$ the set of all polytopes in $\mathbb{R}^n$. 
\end{definition}
\begin{definition}
	(\emph{P-collection}) A P-collection $\mathcal{U}$ is a finite collection of polytopes in $\mathbb{R}^n$, i.e.
	\begin{equation*}
	\mathcal{U} = \cup^{\mathsf{N_c}}_{a=1} \mathcal{P}_a,
	\end{equation*}
	where $\mathsf{N_c}\in \mathbb{N}$, and $\mathcal{P}_a= \{x \in \mathbb{R}^n|P_ax\leq p_a \}$ are polytopes, with $a\in[1,\mathsf{N_c}]$, $P_a\in\mathbb{R}^{n_{p,a} \times n}$, and $p_a\in\mathbb{R}^{n_{p,a}}$.
	Additionally, for a P-collection $\mathcal{U}$, we define
	\begin{equation}
	\mathsf{larg}(\mathcal{U}):= \max_{a\in[1,\mathsf{N_c}]} \mathsf{numh}(\mathcal{P}_a),\label{eq:hyperplane_num}
	\end{equation}
	and 
	\begin{equation}
	\mathsf{num}(\mathcal{U}):= \mathsf{N_c},\label{eq:poly_num}
	\end{equation}
	with $\mathsf{numh}(\cdot)$ as in~\eqref{eq:numh}.
\end{definition}

Now, we formulate the main problem in this work.
\begin{problem}\label{prob}
	Consider a dtLCS $S=(X,X_0,U,W,f)$ as in~\eqref{sys:tuple}, a DSA $\mathcal{A} = (Q, q_0, \Pi,\delta, \text{Acc})$, and a labeling function $L:X\rightarrow \Pi$ as in Definition~\ref{def:L-function}.
	We aim to synthesize a controller (if existing) to enforce the property modeled by $\mathcal{A}$ over $S$.
\end{problem}

For a better illustration of the theoretical results, we also employ a running example throughout this paper.

\begin{example}
	(Running example) 
	Consider a dtLCS as in~\eqref{sys:tuple}, in which
	$A \!=\!  \begin{bmatrix}
	\begin{smallmatrix}0.9990\ &0.1846\\-0.0074\ &0.5265\end{smallmatrix}
	\end{bmatrix}$;
	$B \!=\! \begin{bmatrix}
	\begin{smallmatrix}1.0209;7.3830\end{smallmatrix}
	\end{bmatrix}$;
	$x(k)\!=\![x_1(k);x_2(k)]$ is the state; $X_0=[105,110]\!\times\![-10,10]$ is the initial state set; $u(k)\in[-0.32,\,0.68]$ denotes the input; and $w(k)\in[-0.18,\,0.18]^2$ denotes the disturbances affecting the system.
	Here, we are interested in an $\omega$-regular property $\psi$ which is modeled by a DSA $\mathcal{A}$ as in Figure~\ref{fig:running_dfa}.
	The temporal logic formula\footnote{see~\cite[Section 5.1]{Baier2008Principles} for syntax and semantics of the formula.} for $\mathcal{A}$ is given by $G((p_2\Rightarrow FGp_2)\wedge (\neg p3))$, which, in English, requires that: 
		1) if the system enters the region $X_2: = L^{-1}(p_2)$, it must eventually stay within $X_2$; and 2) the system should not reach the region $X_3:=L^{-1}(p_3)$.
	\begin{figure}[htbp]
		\centering
		\includegraphics[width=0.45\textwidth]{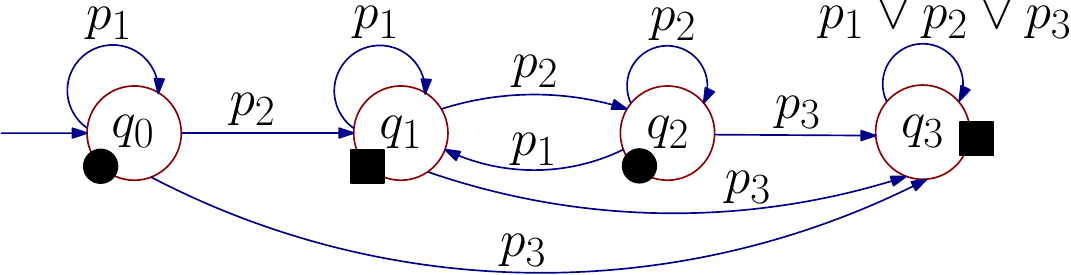}
		\caption{DSA $\mathcal{A}$ modeling $\psi$, with alphabet $\Pi=\{p_1,p_2,p_3\}$; labeling function $L:X \rightarrow \Pi$ with $L(x)=p_1$ when $x \in [105,$ $110]\times[-10,10]$, $L(x)=p_2$ when $x \in (110,115]\times$ $[-10,10]$, and $L(x)=p_3$ when $x \in \mathbb{R}^2\backslash([105, 115]\times[-10,10]) $; and accepting condition $\text{Acc}=\{\langle E_1,F_1\rangle, \langle E_2,F_2\rangle,$ $\langle E_3,F_3\rangle\}$, in which $E_1= \{ q_3\}$, $F_1 = \emptyset$, $E_2 = \{q_1\}$, $F_2 = \{q_2\}$, $E_3 = \emptyset$, and $F_3 = \{q_0\}$. $\blacksquare$ and $\CIRCLE$ indicate the states that can be visited finitely and infinitely many times, respectively. }
		\label{fig:running_dfa}
	\end{figure}
\end{example}

\section{Controller Synthesis via Hybrid Controlled Invariant Sets}\label{sec3}
\subsection{Product System}\label{sec30}
Consider a dtLCS $S$ and a DSA $\mathcal{A}= (Q, q_0, \Pi, \delta, \text{Acc})$.
To solve Problem~\ref{prob}, a product between a dtLCS $S$ and a DSA $\mathcal{A}$ is required, which is formally defined as follows.
\begin{definition}\label{syn_prod}
	\emph{(Product of $S$ and $\mathcal{A}$)}
	Consider a dtLCS $S = (X,X_0,U,W,f)$, a DSA $\mathcal{A}= (Q, q_0, \Pi, $ $\delta, \text{Acc})$,
	and a labeling function $L:X\rightarrow \Pi$.
	The product system between $S$ and $\mathcal{A}$ is defined as
	\begin{equation}\label{f_syn}
	S\otimes\mathcal{A} = (\underline{X}, \underline{X}_0,\underline{U},\underline{W},\underline{f}),
	\end{equation}
	with state set $\underline{X}\!:=\!\{(q,q',x)\!\in\! Q \!\times\! Q\!\times\! X|\exists \sigma\!\in\! \Pi,(q,\sigma,q')\!\in\! \delta,\text{ and }x \!\in\! L^{-1}(\sigma)\}$; 
	the set of initial states $\underline{X}_0:=\{(q_0,q,$ $x)\in\{q_0\}\!\times\! Q\times X_0|\exists \sigma\!\in\! \Pi,(q_0,\sigma,q)\!\in\! \delta,\text{ with }x\! \in \!L^{-1}(\sigma)\}\subseteq\underline{X}$; 
	the input set $\underline{U} := U$;
	and the disturbance set $\underline{W} := W$.
	The transition $\underline{f}:\underline{X}\times\underline{U}\times\underline{W}\rightarrow \underline{X}$ is defined as $\underline{x}':=\underline{f}(\underline{x},u,w)$ with $\underline{x}=(q,q',x),\underline{x}'=(q',q'', x')$, $u\in \underline{U}$, and $w\in \underline{W}$ in which $x' = Ax+Bu+w$ and $(q',L(x'),q'')\in\delta$.
\end{definition}
Consider the hybrid set $\underline{X}$ as in~\eqref{f_syn}, and any set $\underline{X}'\subset \underline{X}$. 
We also need the following definitions in this paper:
\begin{itemize}
	\item \emph{(Projection)} We denote by 
	\begin{align}
	\underline{X}'(q,q') := \{x\in X|(q,q',x)\in \underline{X}'\},\label{proj}
	\end{align}
	the projection of $\underline{X}'$ on $X$ with respect to some $q,q'\in Q$. 
	Accordingly, we define $\big(q,q',\underline{X}'(q,q')\big):=\{(q_1,q_2,x)\in\underline{X}'~|~q_1 = q, q_2 = q'\}$.
	\item \emph{(Hybrid Minkowski sum)} Consider a set $\mathsf{X}\subseteq X$. 
	We denote by $\underline{X}'\oplus\mathsf{X}$ the hybrid Minkowski sum between $\underline{X}'$ and $\mathsf{X}$, which is defined as 
	\begin{align}
	\underline{X}'\oplus\mathsf{X}:=\{(q,q',x)\!\in\!\underline{X}~|~&\underline{X}'(q,q')\neq\emptyset,x\in\underline{X}'(q,q')+\mathsf{X}\};\label{eq:Hms}
	\end{align}
	\item \emph{($\varepsilon$-expansion set)} Consider an $\varepsilon\in\mathbb{R}_{\geq 0}$. 
	We denote by $\underline{X}'_{\varepsilon}$ the $\varepsilon$-expansion of $\underline{X}'$, which is defined as
	\begin{equation}
	\underline{X}'_{\varepsilon} :=\underline{X}'\oplus\varepsilon\mathbb{B}^n; 
	\end{equation}
	\item \emph{($\varepsilon$-contraction set)} Consider an $\varepsilon\in\mathbb{R}_{\geq 0}$.
	We denote by $\underline{X}'_{-\varepsilon}$ the $\varepsilon$-contraction of $\underline{X}'$, which is defined as
	\begin{align}
	\underline{X}'_{-\varepsilon}:=\{(q,q',x)\in\underline{X}&~|~\underline{X}'(q,q')\neq\emptyset, x\in\underline{X}'(q,q')-\varepsilon\mathbb{B}^n\}.\label{eq:vcons}
	\end{align}
	\item \emph{($\rho$-contraction product)} Consider $\rho\in\mathbb{R}_{\geq 0}$. 
	We denote by
	\begin{equation}
	(S\otimes\mathcal{A})_{-\rho} := (\underline{X}_{-\rho}, (\underline{X}_0)_{-\rho},\underline{U}-\rho\mathbb{B}^m,\underline{W},\underline{f}),\label{eq:rho_de_product}
	\end{equation}
	the $\rho$-contraction of $S\otimes\mathcal{A}$ as in~\eqref{f_syn}.
	\item \emph{(Distance)} Consider any $\underline{x}_1,\underline{x}_2\in\underline{X}$, with $\underline{x}_1=(q_1,q'_1,$ $x_1)$ and $\underline{x}_2=(q_2,q'_2,x_2)$.
	The distance between $\underline{x}_1$ and $\underline{x}_2$ is defined as
	\begin{align}
	\!\!\!\!\mathsf{d}(\underline{x}_1,\underline{x}_2)\!:=\!\! \left\lbrace 
	\begin{aligned}
	&+\infty\quad \quad , \text{ if }q_1\!\neq\! q_2\text{ or }q'_1 \!\neq\! q'_2;\\
	&\lVert x_1 - x_2 \rVert, \text{ if }q_1\!=\! q_2\text{ and }q'_1 \!=\! q'_2.
	\end{aligned}
	\right.\label{def:distance}
	\end{align}
\end{itemize}
Additionally, we also define \emph{Hausdorf distance} between any two hybrid sets $\underline{X}',\underline{X}''\subset \underline{X}$ as follows.
\begin{definition}\label{def:hausdorf}
	Consider two hybrid sets $\underline{X}',\underline{X}''\subset \underline{X}$. 
	The \emph{Hausdorf distance} between $\underline{X}'$ and $\underline{X}''$ is defined as
	\begin{align}
	\mathsf{d}_H(\underline{X}',\underline{X}''):= \inf\{\varepsilon\in\mathbb{R}_{\geq 0}|\underline{X}'\!\subseteq\!\underline{X}''_{\varepsilon} \wedge \underline{X}''\!\subseteq\!\underline{X}'_{\varepsilon} \}\label{eq:Hdis}.
	\end{align}
\end{definition}
Next, we proceed with discussing the solution to Problem~\ref{prob}. 
\begin{remark}
	Note that the $\varepsilon$-contraction set in~\eqref{eq:vcons} can be empty when $\varepsilon$ is too large.
	Hence, the $\rho$-contraction products as in~\eqref{eq:rho_de_product} are only meaningful for those $\rho$ with which the sets $\underline{X}_{-\rho}$, $(\underline{X}_0)_{-\rho}$, and $\underline{U}-\rho\mathbb{B}^m$ are not empty.
\end{remark}

\subsection{Synthesis via Hybrid Controlled Invariant Set}\label{sec31}
Here, we show that Problem~\ref{prob} can be solved by computing HCI sets (cf. Definition~\ref{def:HCI}) for the product system as in Definition~\ref{syn_prod}.
To this end, the next result is required.
\begin{theorem}\label{thm:sac}
	Consider a dtLCS $S=(X,X_0,U,W,f)$, a DSA $\mathcal{A} = (Q, q_0, \Pi,\delta, \text{Acc})$, a labeling function $L:X\rightarrow \Pi$, the product $S\otimes\mathcal{A}$ as in Definition~\ref{syn_prod}, 
	and a set $\underline{E}\!\subset\! \underline{X}$ such that $\underline{X}\backslash\underline{E}$ is the state set of the product system $S\otimes \mathcal{A}_{rd}(E')$, with $\mathcal{A}_{rd}(E') := (Q_{rd}, q_0, \Pi_{rd},\delta_{rd},\text{Acc}_{rd})$, and
	\begin{align}
	E':=\{q\in Q | \exists r\in\{1,\ldots,\mathsf{r}\}, q\in E_r \}.\label{E'}
	\end{align}
	\noindent
	One has $S\models \mathcal{A}$ if for any infinite state sequence $\underline{\xi}_{x} = (\underline{x}(0),$ $\underline{x}(1),\ldots,\underline{x}(k),\ldots)$ of $S\otimes\mathcal{A}$, $\underline{x}(k)\notin \underline{E}$, $\forall k\in\mathbb{N}$. 
\end{theorem}
One can show Theorem~\ref{thm:sac} by considering the accepting condition of $\mathcal{A}$ as in~\eqref{acc_cond}.
As a key insight, if one can find a controller that keeps all infinite state sequences of $S\otimes \mathcal{A}$ evolving within the set $\underline{X}\backslash\underline{E}$, then any state $q\in E'$ would be visited \emph{at most once} considering the definition of the reduced DSA $\mathcal{A}_{rd}(E')$.
One can build such a controller by leveraging HCI sets for $S\otimes\mathcal{A}$, as defined next.
\begin{definition}\label{def:HCI}
	(\emph{HCI Set})
	A set $\underline{I}\subseteq\underline{X}\backslash\underline{E}$ is an HCI set for $S\otimes \mathcal{A}$, if $\forall \underline{x}\in \underline{I}$, $\exists u \in \underline{U}$ such that $\forall w\in\underline{W}$,  one has $\underline{x}':=\underline{f}(\underline{x},u,w) \in \underline{I}$, with $\underline{E}$ being the set as in Definition~\ref{thm:sac}.
	Additionally, we denote by $\underline{I}^*$ the \emph{maximal HCI set} in the sense that for any other HCI set $\underline{I}'\subset \underline{X}\backslash\underline{E}$, we have $\underline{I}'\subset \underline{I}^*$.
\end{definition}
Note that the HCI set defined here is similar to the~\emph{strongly reachable set} in~\cite[Definition 2]{Bertsekas1972Infinite}, but defined on the hybrid set $\underline{X}$ instead of $\mathbb{R}^n$.
Based on the definition for the HCI set, we define an \emph{HCI-based controller} as follows.
\begin{definition}\label{def:HCIcontroller}
	(\emph{HCI-based controller})
	Consider a dtLCS $S=(X,X_0,U,W,f)$, a DSA $\mathcal{A} = (Q, q_0, \Pi,\delta, \text{Acc})$, a labeling function $L:X\rightarrow \Pi$, the product system $S\otimes\mathcal{A}$ as in Definition~\ref{syn_prod}, and a non-empty HCI set $\underline{I}$ for $S\otimes \mathcal{A}$.
	An HCI-based controller 
	$\mu:\underline{X}\rightarrow \underline{U}$ is constructed as follows: given $\underline{x}(k) = (q,q',x)$, input $u(k)=\mu(\underline{x}(k))$ should be chosen such that
	$\forall x'\in Ax(k)+Bu(k)+W$, one gets $(q',q'',x')\in\underline{I}$, with $(q',\sigma,q'')\in \delta$ and $\sigma = L(x')$.
\end{definition}
With Definition~\ref{def:HCIcontroller} in hand, the next result shows that once there exists a non-empty HCI set $\underline{I}$, the construction of an HCI-based controller is always feasible.
\begin{proposition}\label{lem:stationary_policy}
	Consider a dtLCS $S$, a DSA $\mathcal{A}$ modeling the desired $\omega$-regular property, and the product system $S\otimes\mathcal{A}$ as in Definition~\ref{syn_prod}. 
	For any non-empty HCI set $\underline{I}$ of $S\otimes\mathcal{A}$, there exists an HCI-based controller $\mu$ as in Definition~\ref{def:HCIcontroller}.  
\end{proposition}
\vspace{0.1cm}
The proof of Proposition~\ref{lem:stationary_policy} is shown in Appendix~\ref{proof:section3}.
By virtue of Definition~\ref{def:HCIcontroller} and Proposition~\ref{lem:stationary_policy}, we reduce Problem~\ref{prob} to the computation of (maximal) HCI sets for $S\otimes \mathcal{A}$. 
In Section~\ref{sec:comHCI}, we discuss how to compute such sets.
\addtocounter{example}{-1}
\begin{example}[continued]
		(Running example)
		For computing HCI set as in Definition~\ref{def:HCI}, we select
		\begin{align}
		\underline{E}:=\!\!\!\!\!\!\!\!\bigcup_{\forall q'\in\{q_1,q_2\}}\!\!\!\!\!\!\!\!\big(q',q_1\!,\underline{X}(q',q_1)\big)\cup\!\!\! \bigcup_{\forall q'\in Q}\!\!\!\big(q'\!,q_3,\underline{X}(q',q_3)\big),\label{choice-E1}
		\end{align}
		for which the corresponding reduced DSA $\mathcal{A}_{rd}(E')$ is depicted in Figure~\ref{fig:reduced} (left).
		\begin{figure}[htbp]
			\centering
			\subfigure{
				\includegraphics[width=0.30\textwidth]{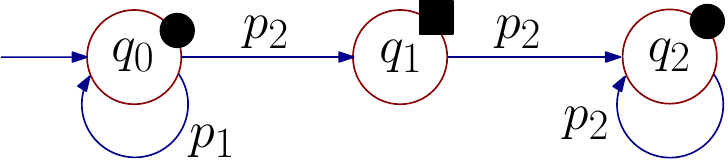}
			}\hspace{0.8cm}
			\quad
			\subfigure{
				\includegraphics[width=0.08\textwidth]{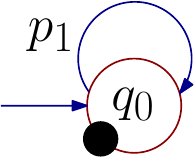}
			}\hspace{-0.4cm}
			\caption{Reduced DSA $\mathcal{A}_{rd}(E')$ for different choice of $\underline{E}$.}
			\label{fig:reduced}
		\end{figure}
		Note that the selection of the set $\underline{E}$ is not unique.
		One can also choose $\underline{E}$ such that $\underline{X}\backslash\underline{E}= \big(q_0,q_0,\underline{X}(q_0,q_0)\big)$, with the underlying reduced DSA as in Figure~\ref{fig:reduced} (right).
		However, such a choice essentially prevents all the states in the set $E'$ as in~\eqref{E'} from being reached, which is more conservative than the choice in~\eqref{choice-E1} (cf. Remark~\ref{re:improve}).
\end{example}
\begin{figure}[htbp]
	\centering
	\subfigure{
		\includegraphics[width=0.40\textwidth]{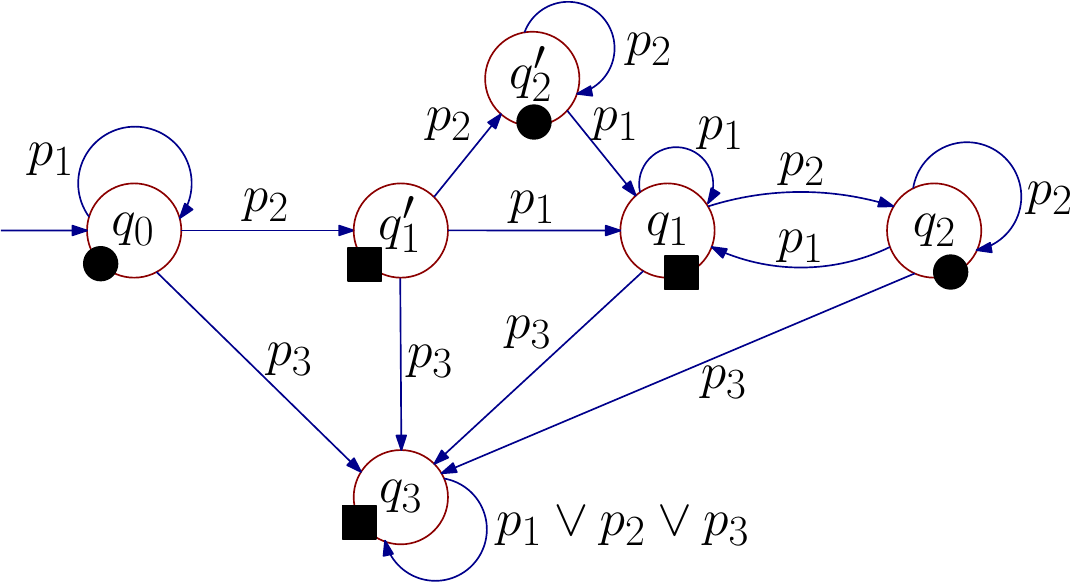}
	}\hspace{-0.7cm}
	\quad
	\subfigure{
		\includegraphics[width=0.20\textwidth]{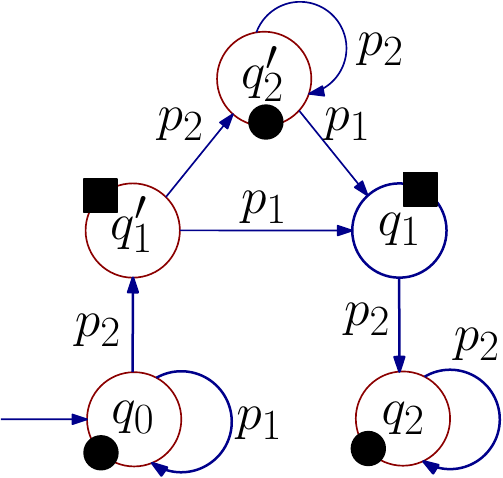}
	}\hspace{-0.4cm}
	\caption{\textbf{Left}: DSA $\mathcal{A}'$ modeling $\psi$, with the same alphabet and labeling function as $\mathcal{A}$ in Figure~\ref{fig:running_dfa}, and accepting condition $\text{Acc}\!=\!\{\langle E_1,F_1\rangle,\langle E_2,F_2\rangle,\langle E_3,F_3\rangle\}$, with $E_1\!=\! \{ q_3\}$, $F_1 \!=\! \emptyset$, $E_2 \!=\! \{q_1,q'_1\}$, $F_2\! =\! \{q_2,q'_2\}$, $E_3 \!=\! \emptyset$, and $F_3 \!=\! \{q_0\}$. Transition $(q'_2,p_3,q_3)$ is omitted to keep the figure less crowded. \textbf{Right}: The reduced DSA of $\mathcal{A}'$ with $\underline{E}$ selected as in~\eqref{choice-E1}. }
	\label{fig:running_dfa_extend}
\end{figure}
\begin{remark}\label{re:improve}
	Theorem~\ref{thm:sac} generalizes the results in~\cite[Theorem 3.2]{Zhong2022Compositional}, since~\cite[eq. (3.7)]{Zhong2022Compositional} essentially provides a special choice of the set $\underline{E}$ in Theorem~\ref{thm:sac} that \emph{prevents states in the set $E'$ in~\eqref{E'} from being reached}.
		It is also worth mentioning that the results in Theorem~\ref{thm:sac} can readily be applied to synthesize controllers that allow some states in $E'$ being visited \emph{at most $N'$ times}, where $N'\in\mathbb{N}_{\geq 1}$ is chosen by the users. 
		For instance, to synthesize a controller that allows $E_2$ of $\mathcal{A}$ being visited at most twice (i.e. $N'=2$), one can first reformulate $\mathcal{A}$ in Figure~\ref{fig:running_dfa} to another DSA $\mathcal{A}'$ as in Figure~\ref{fig:running_dfa_extend} (Left).
		Then, one can apply Theorem~\ref{thm:sac} to $\mathcal{A}'$ by selecting $\underline{E}$ as in~\eqref{choice-E1}, which corresponds to a reduced DSA as in Figure~\ref{fig:running_dfa_extend} (Right),
		and design an HCI-based controller accordingly (if existing).
		For the sake of simple presentation, the formal definition of such reformulation is omitted here.  
		In future work, we plan to work on building controllers that can enforce $\omega$-regular properties by ensuring that for some $r \in\{1,\ldots,\mathsf{r}\}$, $q\in F_r$ are visited infinitely many often so that  enforcing $inf(\mathbf{q})\,\wedge\,E_r=\emptyset$ for these $r$ is not required.
		However, this is beyond the scope of the current work.
\end{remark}

\subsection{Computation of Maximal HCI Set}\label{sec:comHCI}
Inspired by the method proposed in~\cite{Bertsekas1972Infinite} for computing maximal strongly reachable set, we propose the following approach to compute the maximal HCI set. 
\begin{definition}\label{def:iteration}
	Consider a dtLCS $S$ as in~\eqref{sys:tuple}, a DSA $\mathcal{A}$ modeling the desired $\omega$-regular property, the product system $S\otimes\mathcal{A} = (\underline{X}, \underline{X}_0,\underline{U},\underline{W},\underline{f})$, and set $\underline{E}\subset \underline{X}$ selected as in Theorem~\ref{thm:sac}.
	The maximal HCI set for $S\otimes\mathcal{A}$ can be computed with iteration~\eqref{iter_p} and stopping criterion~\eqref{stop_r} as: 
	\begin{align}
	\underline{I}_0 = \underline{X}\backslash \underline{E},\ \underline{I}_{i+1} = \underline{I}_0\cap \textbf{P}(\underline{I}_i),\label{iter_p}\\
	\underline{I}_i= \underline{I}_{i+1},\label{stop_r}
	\end{align}
	where 
	\begin{align}
	\textbf{P}(\underline{I}) = \{\underline{x}\in \underline{X}~|~\exists u \in \underline{U},\forall w\in\underline{W}, \text{ such that }\underline{f}(\underline{x},u,w)\in \underline{I}\},\label{eq:iter_pf}
	\end{align}
	denotes the set of states that reach $\underline{I}$ in one step.
	Once the iteration in~\eqref{iter_p} is terminated by the stopping criterion in~\eqref{stop_r}, $\underline{I}_i$ is the maximal HCI set. 
\end{definition}
To ensure the convergence of the iteration scheme in Definition~\ref{def:iteration}, we have the following assumption.
\begin{assumption}\label{assum1}
	Consider a dtLCS $S$, a DSA  $\mathcal{A}$ representing the desired $\omega$-regular property, a labeling function $L:X\rightarrow \Pi$ as in Definition~\ref{def:L-function}, and the corresponding product system $S\otimes\mathcal{A} = (\underline{X}, \underline{X}_0,\underline{U},\underline{W},\underline{f})$ as in~\eqref{f_syn}.
	We assume:
	\begin{enumerate}
		\item Input set $\underline{U}$ and disturbance set $\underline{W}$ are of the form of polytopes in $\mathbb{R}^m$ and  $\mathbb{R}^n$, respectively;
		\item The set $(\underline{X}\backslash\underline{E})(q,q')$, as defined in~\eqref{proj}, is compact and of the form of a P-collection in $\mathbb{R}^n$, $\forall q,q'\!\in\! Q$.
	\end{enumerate}
\end{assumption}
\begin{algorithm}
	\caption{Computing maximal HCI Set $\underline{I}^*$ } \label{alg}
	\label{alg:Framwork} 
	\KwIn{$\underline{X}\backslash\underline{E}$, $S\otimes \mathcal{A}$}
	\KwOut{Maximal HCI set $\underline{I}^*$}
	$i=0$, $\underline{I}_0=\underline{X}\backslash\underline{E}$\label{start}\\
	\While{$1$}
	{
		$\underline{I}_{i+1}=\emptyset$, $Pr = \emptyset$;\label{Pr_start} \\
		\For{every $(q',q'')$ s.t. $\exists x$, $(q',q'',x)\in \underline{I}_i$}{
			$Proj = pre(\underline{I}_i(q',q''))$;\label{state_f_proj}\\
			\For{every $q\in Q$ s.t. $\exists \sigma\in \Pi$, $(q,\sigma,q')\in\delta$}{
				$Pr=Pr\,\cup\,\{(q,q',x)|x\in Proj\}$;\label{Pr} 
			} 
		}
		\For{every $(q,q')$ s.t. $\exists x$, $(q,q',x)\in Pr$}{\label{int1}
			$I_c = \underline{I}_0(q,q')\cap Pr(q,q')$;\label{int2}\\
			$\underline{I}_{i+1}=\underline{I}_{i+1}\,\cup\,\{(q,q',x)|x\in I_c\};$ \label{int3}
		}
		\uIf{$\underline{I}_i= \underline{I}_{i+1}$}{\label{sucess_start}
			$\underline{I}^*=\underline{I}_i$;\\
			Stop successfully; \label{sucess_end}
		}
		\uElseIf{$\underline{I}_{i+1}$ is empty}{\label{unsucess_start}
			Stop unsuccessfully;\label{unsucess_end}
		}
		\Else{
			$i=i+1$;
		}
	}
\end{algorithm}
\setlength{\textfloatsep}{0pt}
With Definition~\ref{def:iteration} and Assumption~\ref{assum1}, we show that $\underline{I}_i$ converges to maximal HCI set $\underline{I}^*$ as $i$ goes to infinity.
\begin{theorem}\label{thm:convergence}
	Consider a dtLCS $S$ as in Definition~\ref{def:dtLCS}, and a DSA $\mathcal{A}$ modeling the desired $\omega$-regular property such that Assumption~\ref{assum1} holds.
	Then, considering the iteration in~\eqref{iter_p}, we have $\underline{I}^* = \lim\limits_{i\rightarrow \infty} \underline{I}_i$,
	where the limit is in terms of the Hausdorff distance as in Definition~\ref{def:hausdorf}.
\end{theorem}
The proof of Theorem~\ref{thm:convergence} is inspired by~\cite{Bertsekas1972Infinite} and can be found in Appendix~\ref{proof:section3}.
Next, we discuss the implementation of~\eqref{iter_p} and~\eqref{stop_r}.
Considering the dynamics as in~\eqref{eq:linear_subsys}, by the definition of $\underline{f}$, $\textbf{P}(\underline{I})$ as in~\eqref{eq:iter_pf} can be rewritten as
\begin{align}
\textbf{P}(\underline{I})  = \{ (q,q',x)\in \underline{X}\,|\, x\in pre(\underline{I}(q',q'')),\text{ with } q,q',q''\in Q\text{ s.t. }\exists\sigma\in\Pi, (q,\sigma,q')\in \delta\},\label{eq:pr}
\end{align}
with 
\begin{align}
pre(X')=\{x\in X|\exists&u\in U, \ \forall w \in W,\  Ax+Bu+w\in X'\},\label{eq:pre_orig}
\end{align}
and $\underline{I}(q',q'')\neq \emptyset$ as defined in~\eqref{proj}.
Informally, $pre(X')$ computes the \emph{one-step-backward projection} of the set $X'$ considering the linear dynamics as in~\eqref{eq:linear_subsys}.
Based on~\eqref{eq:pr}, we present the main implementation in Algorithm~\ref{alg}.
In each iteration, $\textbf{P}(\underline{I}_i)$ is computed as in line~\ref{Pr_start}-\ref{Pr}, where line~\ref{Pr} and~\ref{state_f_proj} correspond to~\eqref{eq:pr} and~\eqref{eq:pre_orig}, respectively; 
	$\underline{I}_0\cap \textbf{P}(\underline{I}_i)$ is computed as in line~\ref{int1}-\ref{int3}.
	In particular, one can readily employ existing toolboxes, including multi-parametric toolbox \texttt{MPT}~\cite{Herceg2013Multia} and \texttt{BENSOLVE}~\cite{Loehne2016Equivalence}, to perform those polyhedral operations in each iteration. 
	The iteration proceeds until either: 1) $\underline{I}_i= \underline{I}_{i+1}$ (line~\ref{sucess_start}-\ref{sucess_end}); or 2) $\underline{I}_{i+1}=\emptyset$ (line~\ref{unsucess_start}-\ref{unsucess_end}), meaning a non-empty HCI set does not exist.
\begin{remark}\label{deflate}
	If the set $\underline{X}\backslash\underline{E}$ is not compact, one can reselect the set $\underline{E}$ to ensure (if possible) the compactness of $\underline{X}\backslash\underline{E}$.
		Additionally, one can also (slightly) deflate the original set $\underline{X}\backslash\underline{E}$ such that one can start Algorithm~\ref{alg} with a compact $\underline{X}\backslash\underline{E}$.
		Such deflation is shown using the running example.
\end{remark}
\addtocounter{example}{-1}
\begin{example}[continued]
	\begin{figure}[htbp]
		\centering
		\subfigure{
			\includegraphics[width=0.55\textwidth]{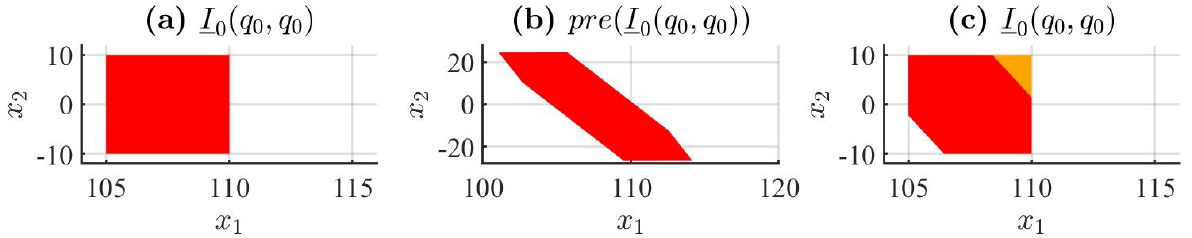}
		}\hspace{-0.4cm}
		\quad
		\subfigure{
			\includegraphics[width=0.55\textwidth]{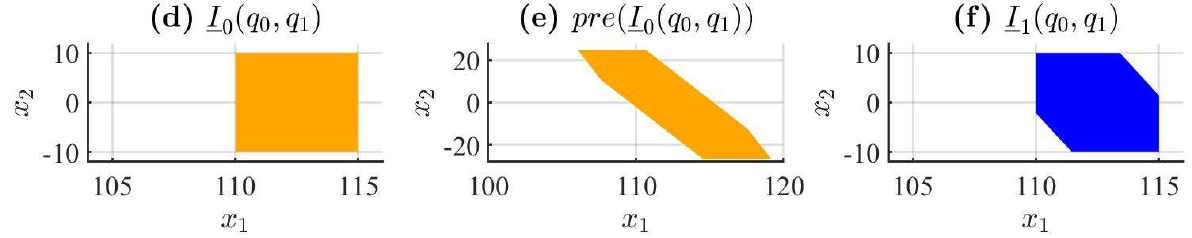}
		}\hspace{-0.4cm}
		\quad
		\subfigure{ 
			\includegraphics[width=0.55\textwidth]{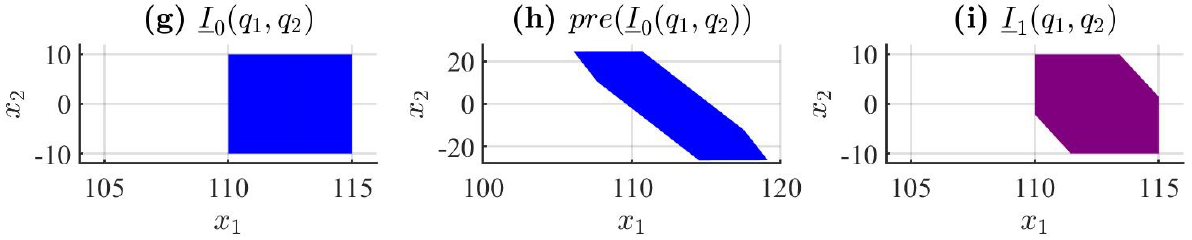}
		}
		\quad
		\subfigure{ 
			\includegraphics[width=0.55\textwidth]{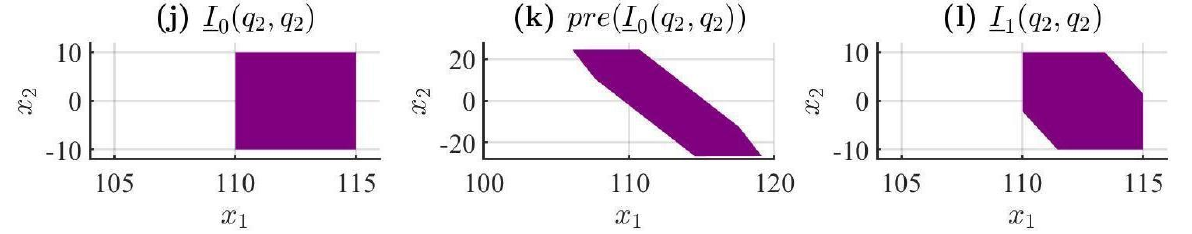}
		}
		\caption{Computation of $\underline{I}_1$ based on $\underline{I}_0$ for the running example according to Algorithm~\ref{alg}.}
		\label{fig:norminal_iter}
	\end{figure}
	(Running example) 
	With $\underline{E}$ selected as in~\eqref{choice-E1}, the set $\underline{X}\backslash\underline{E}$ is not compact.
		Nevertheless, following the idea of Remark~\ref{deflate}, one can ensure the compactness of $\underline{X}\backslash\underline{E}$ by slightly deflating it such that
		$\underline{X}\backslash\underline{E}(q_0,q_1)=\underline{X}\backslash\underline{E}(q_1,q_2) = [110+\epsilon,115]\times[-10,10]$, with $\epsilon\in \mathbb{R}_{>0}$ being any arbitrary positive real number.
		Here, we select $\epsilon=0.01$ and proceed with the computation as in Algorithm~\ref{alg}.
	To provide more intuition on how Algorithm~\ref{alg} works, we demonstrate in Figure~\ref{fig:norminal_iter} the computation of $\underline{I}_1$ based on $\underline{I}_0$ for the running example.
	Concretely, the iteration starts from $\underline{I}_0$ as depicted in Figure~\ref{fig:norminal_iter}($a$), ($d$), ($g$), and ($j$) (cf. line~\ref{start} in Algorithm~\ref{alg}).
	Then, by leveraging~\eqref{eq:pre_orig}, we compute the one-step-backward projection of $\underline{I}_0(q_0,q_0)$, $\underline{I}_0(q_0,q_1)$, $\underline{I}_0(q_1,q_2)$, and $\underline{I}_0(q_2,q_2)$, as shown in Figure~\ref{fig:norminal_iter}($b$), ($e$), ($h$), and ($k$) , respectively (cf. line~\ref{state_f_proj} in Algorithm~\ref{alg}).
	Based on these projections, $\mathbf{P}(\underline{I}_0)$ as in~\eqref{eq:pr} are computed (cf. line~\ref{Pr} in Algorithm~\ref{alg}), in which
	\begin{align*}
		\mathbf{P}(\underline{I}_0)(q_0,q_0) &= \big(q_0,q_0,pre(\underline{I}_0(q_0,q_0) \big);\\
		\mathbf{P}(\underline{I}_0)(q_0,q_1) &= \big(q_0,q_0,pre(\underline{I}_0(q_0,q_1) \big);\\
		\mathbf{P}(\underline{I}_0)(q_1,q_2) &= \big(q_0,q_1,pre(\underline{I}_0(q_1,q_2) \big);\\
		\mathbf{P}(\underline{I}_0)(q_2,q_2) &= \big(q_1,q_2,pre(\underline{I}_0(q_2,q_2) \big)\cup \big(q_2,q_2,pre(\underline{I}_0(q_2,q_2) \big).
		\end{align*}
	Finally, we compute $\underline{I}_1$ as in~\eqref{iter_p} based on $\mathbf{P}(\underline{I}_0)$ (cf. line~\ref{int2} to~\ref{int3} in Algorithm~\ref{alg}).
	Accordingly, one obtains 
	\begin{align*}
		\underline{I}_1= \ &\big(q_0,q_0, \underline{I}_1(q_0,q_0) \big)\cup \big(q_0,q_1, \underline{I}_1(q_0,q_1)\big)\cup \big(q_1,q_2, \underline{I}_1(q_1,q_2)\big)\cup \big(q_2,q_2, \underline{I}_1(q_2,q_2) \big),
		\end{align*}
	in which
	\begin{align*}
		\underline{I}_1(q_0,q_0) &= \underline{I}_0(q_0,q_0) \cap \big(pre(\underline{I}_0(q_0,q_0))\cup pre(\underline{I}_0(q_0,q_1))\big);\\
		\underline{I}_1(q_0,q_1) &= \underline{I}_0(q_0,q_1) \cap pre(\underline{I}_0(q_1,q_2));\\
		\underline{I}_1(q_1,q_2) &= \underline{I}_0(q_1,q_2) \cap pre(\underline{I}_0(q_2,q_2)),\\
		\underline{I}_1(q_2,q_2) &= \underline{I}_0(q_2,q_2) \cap pre(\underline{I}_0(q_2,q_2)),
		\end{align*}
	\noindent
	as illustrated in Figure~\ref{fig:norminal_iter} ($c$), ($f$), ($i$), and ($l$), respectively.
\end{example}
It is worth mentioning that when invariance properties are of interest, the iteration in~\eqref{iter_p} terminates within a finite number of steps if there are additional assumptions on the system dynamics (see e.g.~\cite[Proposition 4]{Vidal2000Controlled}), or if $X$, $U$, and $W$ have special shapes (see e.g.~\cite[Theorem 3.1]{Gutman1986Admissible},~\cite[Theorem 5]{Vidal2000Controlled},~\cite[Proposition 5.9]{Blanchini2015Set},\cite[Theorem 1 and Corollary 1]{Rakovic2004Computation}).
However, there is no guarantee that~\eqref{iter_p} can be terminated within a finite number of iterations, in general.
This issue motivates us to propose two alternative iterative schemes to compute approximations of $\underline{I}^*$ (if existing) within a finite number of iterations, which are introduced in Section~\ref{sec4}.

\section{Approximation of Maximal HCI Sets}\label{sec4}
In this section, we propose two methods for computing approximations of $\underline{I}^*$ for $S\otimes\mathcal{A}$ within a finite number of iterations.
For both methods, the following assumption for the dtLCS is required.
\begin{assumption}\label{asum:ctrb}
	Consider a dtLCS $S$ as in Definition~\ref{def:dtLCS}.
	We assume that ($A$,$B$) in~\eqref{eq:linear_subsys} is controllable.
\end{assumption}

\subsection{($\varepsilon_x$,$\varepsilon_u$)-Contraction-based Approximation}
In this subsection, we show how to compute an \emph{($\varepsilon_x$,$\varepsilon_u$)-contraction-based approximation} of $\underline{I}^*$ for $S\otimes\mathcal{A}$.
This approximation is computed based on \emph{a sequence of ($\varepsilon_x,\varepsilon_u$)-constraint $i$-step null-controllable sets}, as defined below.
\begin{definition}\label{def:seq:Ni}
	Consider a dtLCS $S$ as in Definition~\ref{def:dtLCS} in which $W = \{\mathbf{0}_n\}$, and some $\varepsilon_x,\varepsilon_u\in\mathbb{R}_{>0}$.
	A sequence of \emph{($\varepsilon_x,\varepsilon_u$)-constraint $i$-step null-controllable sets}, denoted by $(\mathcal{N}_i(\varepsilon_x,\varepsilon_u))_{i\in\mathbb{N}}$, is recursively defined as 
	\begin{align}
	\mathcal{N}_0(\varepsilon_x,\varepsilon_u) =&\{\mathbf{0}_n\},\nonumber\\
	\mathcal{N}_{i+1}(\varepsilon_x,\varepsilon_u) =& \{x\in\mathbb{R}^n|\exists u\in\varepsilon_u \mathbb{B}^m, Ax +Bu\in \mathcal{N}_i(\varepsilon_x,\varepsilon_u)\}\cap \varepsilon_x \mathbb{B}^n.~\label{eq:seq}
	\end{align}
\end{definition}
Moreover, we have the following lemma for $(\mathcal{N}_i(\varepsilon_x,\varepsilon_u))_{i\in\mathbb{N}}$.
\begin{lemma}\label{lem:cxcu}
	Consider a dtLCS $S$ as in~\eqref{eq:linear_subsys} in which $W =$ $\{\mathbf{0}_n\}$ and ($A$,$B$) is controllable, and a sequence $(\mathcal{N}_i(\varepsilon_x,\varepsilon_u))_{i\in\mathbb{N}}$ as defined in~\eqref{eq:seq}.
	Then, $\exists c_x,c_u\in\mathbb{R}_{>0}$ and $n'\in\mathbb{N}$ with $n'\leq n$ such that $\forall \gamma\in\mathbb{R}_{>0}$, one has
	\begin{equation}
	\gamma\mathbb{B}^n\subseteq \mathcal{N}_{n'}(\varepsilon_x,\varepsilon_u),\label{eq:lem:cxcu}
	\end{equation}
	with $\varepsilon_x=c_x\gamma$ and $\varepsilon_u=c_u\gamma$.
\end{lemma}
The proof of Lemma~\ref{lem:cxcu} is inspired by~\cite[Lemma 2]{Rungger2017Computing} and given in Appendix~\ref{proof:section4.1}.
Note that $c_x$, $c_u$, and $n'$ in Lemma~\ref{lem:cxcu} can be obtained by leveraging the next Corollary.
\begin{corollary}\label{col:cxcu}
	Consider the vertices $z_i\in\mathbb{R}^n$ of $\mathbb{B}^n$, with $i\in[1,2^n]$.
	One can select any $c_x, c_u\in\mathbb{R}^n$, and $n'\in\mathbb{N}$ for Lemma~\ref{lem:cxcu} such that~\eqref{eq:lem:cxcu} holds, if the following constraints are respected for all $z_i$:
	\begin{align}
	&A^{n'}z_i+\sum_{j=0}^{n'-1}A^{n'-j-1}Bu_j=\mathbf{0}_n;\label{eq:cd1}\\
	&|u_j|\leq c_u,\,\forall j\in[0,n'-1];\label{eq:cd2}\\
	&|A^dz_i+\sum_{j=0}^{d-1}A^{d-j-1}Bu_j|\leq c_x,\forall d\in[1,n'-1],\label{eq:cd3}
	\end{align}
	with $u_j\in\mathbb{R}^m$, $j\in[0,n'-1]$. 
\end{corollary}
The proof of Corollary~\ref{col:cxcu} is provided in Appendix~\ref{proof:section4.1}.
Next, we propose the computation of ($\varepsilon_x$,$\varepsilon_u$)-contraction-based approximation in Definition~\ref{def:outap}.
\begin{definition}\label{def:outap}
	\emph{(($\varepsilon_x$,$\varepsilon_u$)-contraction-based approximation)}
	Consider a dtLCS $S$ as in Definition~\ref{def:dtLCS} such that Assumption~\ref{asum:ctrb} holds, a DSA $\mathcal{A}$ modeling the desired property, and the product system $S\otimes\mathcal{A} = (\underline{X}, \underline{X}_0,\underline{U},\underline{W},\underline{f})$. 
	Given $c_x,c_u\in\mathbb{R}_{>0}$ as in Corollary~\ref{col:cxcu}, and any $\gamma\in\mathbb{R}$, we define iteration~\eqref{iter_po} and stopping criterion~\eqref{stop_ro} for computing the ($\varepsilon_x$,$\varepsilon_u$)-contraction-based approximation as:
	\begin{align}
	\underline{I}_0 = (\underline{X}\backslash \underline{E})_{-\varepsilon_x},\ \underline{I}_{i+1} = \underline{I}_0\cap \textbf{P}_{(\varepsilon_x,\varepsilon_u)}(\underline{I}_i),\label{iter_po}\\
	\underline{I}_i\subseteq (\underline{I}_{i+n'})_{\gamma},\label{stop_ro}
	\end{align}
	where  $\varepsilon_x$, $\varepsilon_u$, and $n'$ are as in Lemma~\ref{lem:cxcu} s.t.~\eqref{eq:lem:cxcu} holds, $\textbf{P}_{(\varepsilon_x,\varepsilon_u)}(\underline{I})$ is defined similarly to $\textbf{P}(\underline{I})$ as in~\eqref{eq:pr}, with 
	\begin{align}
	pre(X')=\{x\in X|\exists u\in U-\varepsilon_u\mathbb{B}^m , \forall w \in W, Ax+Bu+w \in X'\}.\label{eq:pre_cont}
	\end{align}
\end{definition}
By leveraging the iteration and stopping criterion as in Definition~\ref{def:outap}, we are able to construct the ($\varepsilon_x$,$\varepsilon_u$)-contraction-based approximation using the following result.
\begin{theorem}\label{thm:c-basedapp}
	For any $\gamma\in\mathbb{R}_{>0}$ and the corresponding $\varepsilon_x,\varepsilon_u\in\mathbb{R}_{>0}$, $n'\in\mathbb{N}$ as in Lemma~\ref{lem:cxcu}, there exists $i\in\mathbb{N}$ with which~\eqref{stop_ro} holds.
	Moreover, consider $(\underline{I}_i)_{i\in\mathbb{N}}$ that is obtained through the iteration as in~\eqref{iter_po}, and the sequence $(\mathcal{N}_i(\varepsilon_x,\varepsilon_u))_{i\in\mathbb{N}}$ as in Definition~\ref{def:seq:Ni}.
	The set
	\begin{align}\label{eq:Icbased}
	\underline{I}(\varepsilon_x,\varepsilon_u) = \bigcup_{i'\in [1,n']} (\underline{I}_{i_*+i'}\oplus \mathcal{N}_{i'}(\varepsilon_x,\varepsilon_u)),
	\end{align}
	is an HCI set for the product system $S\otimes\mathcal{A}$, with $i^*\in\mathbb{N}$ being the smallest index $i$ for the given $\gamma$ such that~\eqref{stop_ro} holds.
\end{theorem}
The proof of Theorem~\ref{thm:c-basedapp} can be found in Appendix~\ref{proof:section4.2}.
Note that the existence of $i\in\mathbb{N}$ such that~\eqref{stop_ro} holds indicates that the iteration in~\eqref{iter_po} can be terminated within finite number of iterations.
Since $\underline{I}(\varepsilon_x,\varepsilon_u)$ in~\eqref{eq:Icbased} is an HCI set for $S\otimes\mathcal{A}$, it is, by definition, an under-approximation of the maximal HCI set $\underline{I}^*$ according to Definition~\ref{def:HCI}.
With the next result, we show how close this approximation is.
In brief, we show that given a $\rho\in\mathbb{R}_{>0}$ and a product system $(S\otimes\mathcal{A})_{-\rho}$ as defined in~\eqref{eq:rho_de_product}, we are able to construct an ($\varepsilon_x$,$\varepsilon_u$)-contraction-based approximation that contains the maximal HCI set for $(S\otimes\mathcal{A})_{-\rho}$ by selecting $\varepsilon_x$ and $\varepsilon_u$ properly.
\begin{theorem}\label{thn:cbasedcloss}
	Consider a dtLCS $S$ as in Definition~\ref{def:dtLCS} such that Assumption~\ref{asum:ctrb} holds, a DSA $\mathcal{A}$ modeling the desired property, and the product system $S\otimes\mathcal{A} = (\underline{X}, \underline{X}_0,\underline{U},\underline{W},\underline{f})$.
	For any $\rho\in\mathbb{R}_{>0}$, there exists $\gamma\in\mathbb{R}_{>0}$, such that
	\begin{equation}
	\underline{I}^*_\rho\subseteq \underline{I}(\varepsilon_x,\varepsilon_u),\label{eq:cbasedgood}
	\end{equation}
	where $\underline{I}^*_\rho$ is the maximal HCI set for $(S\otimes\mathcal{A})_{-\rho}$ as defined in~\eqref{eq:rho_de_product}, $\underline{I}(\varepsilon_x,\varepsilon_u)$ is as in~\eqref{eq:Icbased} with $\varepsilon_x$ and $\varepsilon_u$ being computed as in Lemma~\ref{lem:cxcu} based on $\gamma$.
\end{theorem}
The proof of Theorem~\ref{thn:cbasedcloss} is provided in Appendix~\ref{proof:section4.2}.
\begin{figure}[htbp]
	\centering
	\includegraphics[width=0.45\textwidth]{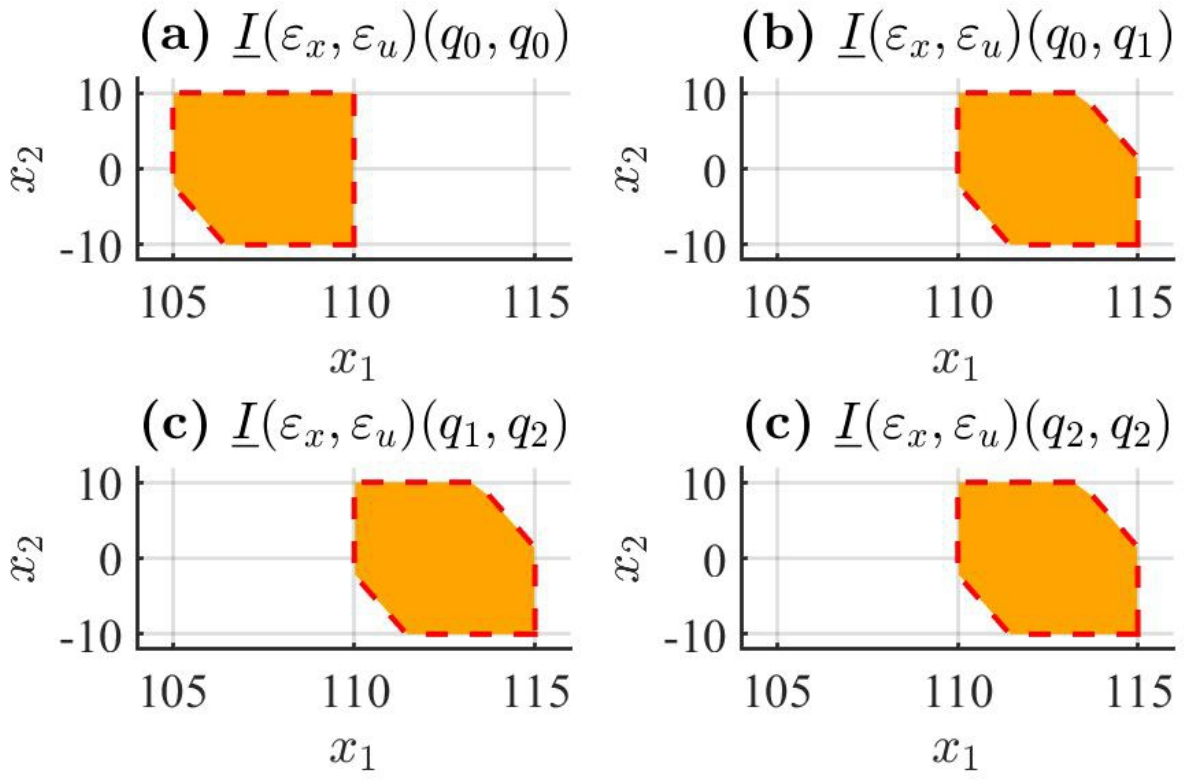}
	\caption{Result for ($\varepsilon_x$,$\varepsilon_u$)-contraction-based approximation (orange region), with $\varepsilon_x=2.8636$ and $\varepsilon_u=0.67251$, and the actual maximal HCI set $\underline{I}^*$ (red dashed lines).}
	\label{fig:result_cbased}
\end{figure}
\addtocounter{example}{-1}
\begin{example}[continued]
	To compute the ($\varepsilon_x$,$\varepsilon_u$)-contraction-based approximation, we choose $n\!=\!2$ and $\gamma\! =\! 0.01$.
	and get $\varepsilon_x\!=\!2.86$ and $\varepsilon_u\!=\!0.67$ considering Lemma~\ref{lem:cxcu} and Corollary~\ref{col:cxcu}.
	Then, we compute the approximation applying Definition~\ref{def:outap} and Theorem~\ref{thm:c-basedapp}.
	The computation ends within 1.36 seconds with 4 iterations.
	The approximation contains 49 hyperplanes, and it is depicted in Figure~\ref{fig:result_cbased}.
	For comparison purposes, we also show the actual maximal HCI set $\underline{I}^*$.
\end{example}

\subsection{$\varepsilon$-Expansion-based Approximation}
Here, we discuss the computation of an $\varepsilon$-expansion-based approximation of the maximal HCI set for $S\otimes\mathcal{A}$.
Such approximations can be computed as in Definition~\ref{def:ebased}.
\begin{definition}\label{def:ebased}
	\emph{($\varepsilon$-expansion-based approximation)}
	Consider a dtLCS $S$ as in~\eqref{eq:linear_subsys} such that Assumption~\ref{asum:ctrb} holds, a DSA $\mathcal{A}$ modeling the desired property, and the product system $S\otimes\mathcal{A} = (\underline{X}, \underline{X}_0,\underline{U},\underline{W},\underline{f})$. 
	Given $\varepsilon\in\mathbb{R}_{>0}$, we define iteration~\eqref{iter_pi} and stopping criterion~\eqref{stop_ri} for computing the $\varepsilon$-expansion-based approximation as:
	\begin{align}
	\underline{I}_0 = \underline{X}\backslash \underline{E},\ \underline{I}_{i+1} = \underline{I}_0\cap \textbf{P}_{\varepsilon}(\underline{I}_i),\label{iter_pi}\\
	\underline{I}_i\subseteq (\underline{I}_{i+1})_{\varepsilon},\label{stop_ri}
	\end{align}
	in which $\textbf{P}_{\varepsilon}(\underline{I})$ is defined similarly to $\textbf{P}(\underline{I})$ as in~\eqref{eq:pr}, with 
	\begin{align}
	pre(X')=\{x\in X|\exists u\  \in U, \forall w \in W',  Ax+Bu+w\in X'\},\label{eq:pre_eb}
	\end{align}
	and $W' := W+\varepsilon\mathbb{B}^n$.
\end{definition}
Unlike~\eqref{eq:pre_orig}, $pre(X')$ as in~\eqref{eq:pre_eb} is defined based on an \emph{$\varepsilon$-expansion} of the set $W$, i.e. $W+\varepsilon\mathbb{B}^n$.
With Definition~\ref{def:ebased}, the next theorem shows the termination of~\eqref{iter_pi} and the construction of the $\varepsilon$-expansion-based approximation. 
\begin{theorem}\label{thm:e-basedapp}
	Consider any $\varepsilon\in\mathbb{R}_{>0}$.
	There exists $i\in\mathbb{N}$ with which~\eqref{stop_ri} holds.
	Additionally, the set 
	\begin{equation}
	\underline{I}(\varepsilon):=\underline{I}_{i^*+1},\label{eq:HCIebased}
	\end{equation}
	is an HCI set for the product system $S\otimes\mathcal{A}$, with $i^*\in\mathbb{N}$ being the smallest index $i$ for the given $\varepsilon$ such that~\eqref{stop_ri} holds.
\end{theorem}
The proof of Theorem~\ref{thm:e-basedapp} can be found in Appendix~\ref{proof:section4.2}.
Note that $\underline{I}(\varepsilon)$ in~\eqref{eq:HCIebased} is an HCI set for $S\otimes\mathcal{A}$, it is therefore also an under-approximation of the maximal HCI set $\underline{I}^*$ according to Definition~\ref{def:HCI}.
Then, similar to Theorem~\ref{thn:cbasedcloss}, we propose the next result to illustrate how close this approximation is.
\begin{theorem}\label{thn:ebasedcloss}
	Consider a dtLCS $S$ as in~\eqref{eq:linear_subsys} such that Assumption~\ref{asum:ctrb} holds, a DSA $\mathcal{A}$ modeling the desired property, and the product system $S\otimes\mathcal{A} = (\underline{X}, \underline{X}_0,\underline{U},\underline{W},\underline{f})$.
	For any $\rho\in\mathbb{R}_{>0}$, there exists $\varepsilon\in\mathbb{R}_{>0}$, such that
	\begin{equation}
	\underline{I}^*_\rho\subseteq \underline{I}(\varepsilon),\label{eq:ebasedgood}
	\end{equation}
	where $\underline{I}^*_\rho$ is the maximal HCI set for $(S\otimes\mathcal{A})_{-\rho}$ as defined in~\eqref{eq:rho_de_product}, and $\underline{I}(\varepsilon)$ is as in~\eqref{eq:HCIebased}.
\end{theorem}
The proof of Theorem~\ref{thn:ebasedcloss} can be found in Appendix~\ref{proof:section4.2}.
\addtocounter{example}{-1}
\begin{example}[continued]
	(Running example) Here, we select $\varepsilon = 0.1$ and compute the $\varepsilon$-expansion-based approximation by applying Definition~\ref{def:ebased} and Theorem~\ref{thm:e-basedapp}.
	The computation terminates within 1.26 seconds with 3 iterations.
	The approximation contains 36 hyperplanes and it is illustrated in Figure~\ref{fig:result_ebased}.
	Additionally, we also depict the actual maximal HCI set $\underline{I}^*$ for comparison purposes.
	\begin{figure}[htbp]
		\centering
		\includegraphics[width=0.45\textwidth]{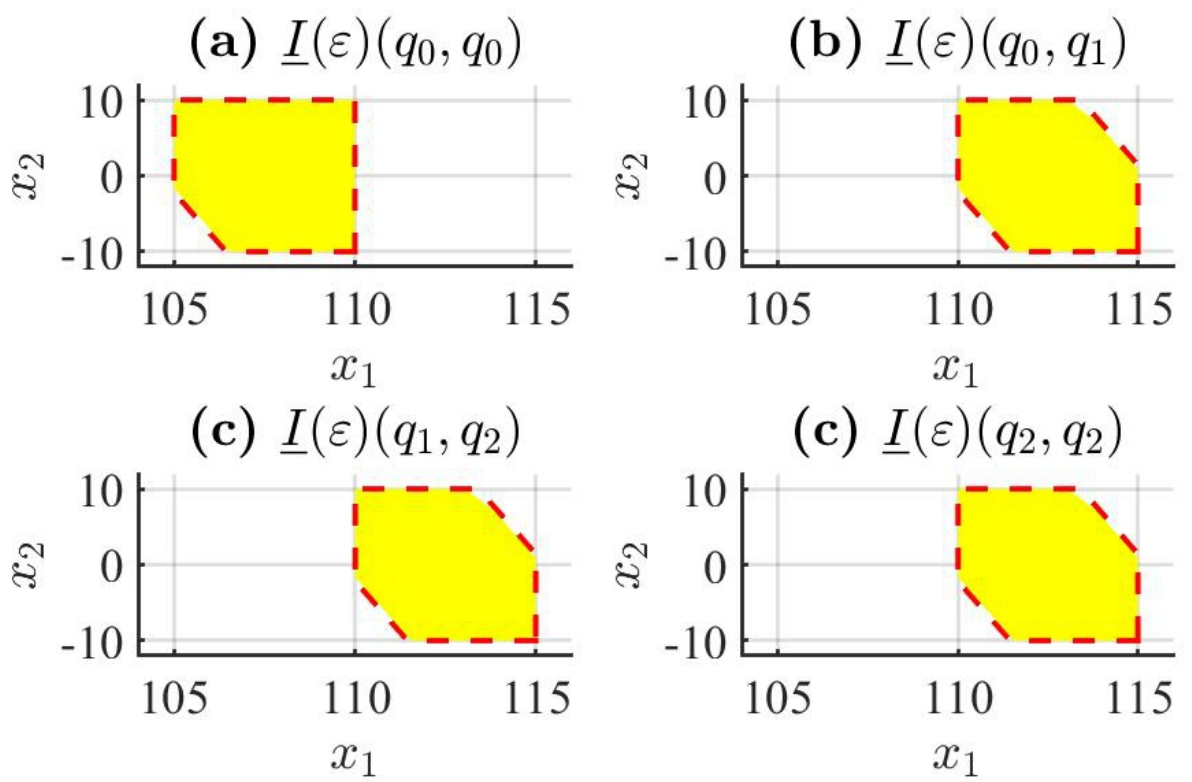}
		\caption{Result for $\varepsilon$-expansion-based approximation (yellow region), with $\varepsilon = 0.1$, and the actual maximal HCI set $\underline{I}^*$ (red dashed lines).}
		\label{fig:result_ebased}
	\end{figure}
\end{example}

\section{Complexity}\label{sec:compl}
In this section, we discuss the space and time complexities of our proposed approaches.
Note that the space and time complexities for the cases in which $W$ has a non-empty interior is still open.
As a key insight, considering a P-collection, denoted by $X':=\cup_{a=1}^{\mathsf{N}_c}X'_a$, one can verify that
\begin{equation}
\mathsf{larg}\Big(pre(X')\Big)=\mathsf{larg}\Big((X'-W)+(-BU)\Big),\label{eq:pre_msum}
\end{equation}
holds by employing the results in~\cite[Section 3.3.3, pp. 44]{Kerrigan2001Robust}, in which $\mathsf{larg}(\cdot)$ is defined in~\eqref{eq:hyperplane_num}, and $BU$ denotes the linear mapping of the input set $U$ regarding matrix $B$~\cite[Section 3.4.2]{Kerrigan2001Robust}. 
However, if $\exists j,k\in[1,\mathsf{N}_c]$ such that $X'_j\cap X'_k\neq \emptyset$, i.e. $X'_a$ are not pairwise disjoint, it is still an open problem for what is the upper bound of the number of polytopes within $X'-W$, and what is the maximal number of hyperplanes defining each polytope within $X'-W$.
Thus, in the remaining discussion, we only focus on the case in which $W= \{\mathbf{0}_n\}$.
To derive the space and time complexities for this case, the following definitions are required.
\begin{definition}\label{helpdef}
	Consider a dtLCS $S\!=\!(X,X_0,U,W,f)$ with $W= \{\mathbf{0}_n\}$, and $p\in \mathbb{N}$.
	We define $\tilde{g}_{S}:\mathbb{N}\rightarrow \mathbb{N}$ as
	\begin{equation}
	\tilde{g}_S(p): = \max_{X'\in\mathcal{P}(n),\text{ with }\mathsf{numh}(X')=p} \mathsf{numh}(pre(X')),\label{gS}
	\end{equation}
	with $\mathsf{numh}(\cdot)$ defined as in~\eqref{eq:numh}, $pre(\cdot)$ defined as in~\eqref{eq:pre_orig}, $n$ being the dimension of $X$, and $X'\subseteq X$.
\end{definition}
\begin{definition}
	Consider a dtLCS $S\!=\!(X,X_0,U,W,f)$ and a DSA $\mathcal{A} = (Q, q_0, \Pi,\delta, \text{Acc})$ modeling the desired $\omega$-regular property.
		We define the set
		\begin{align}
		Q_{rd} :=\{ q\in Q | \exists q' \in Q, \text{ such that } \underline{X}\backslash\underline{E}(q',q)\ne \emptyset\text{ or } \underline{X}\backslash\underline{E}(q,q')\ne \emptyset\},\label{Qrd}
		\end{align}
		with the set $\underline{E}$ being defined in Theorem~\ref{thm:sac}. 
\end{definition}
Intuitively, $\tilde{g}_S(p)$ denotes the maximal number of hyperplanes defining $pre(X')$, with $X'$ being any arbitrary polytope defined by $p$ hyperplanes. 
The set $Q_{rd}$ is the finite state set of the reduced DSA corresponding to the set $\underline{E}$.
With these definitions, we propose the next result that paves the way for deriving the worst-case space and time complexities.
\begin{theorem}\label{thm:upperbound}
	Consider a dtLCS $S=(X,X_0,U,W,f)$ with $W= \{\mathbf{0}_n\}$, a DSA $\mathcal{A} = (Q, q_0, \Pi,\delta, \text{Acc})$ modeling the desired $\omega$-regular property, and the sequence of $\underline{I}_i$ with $i\in\mathbb{N}$ as defined in~\eqref{iter_p} and~\eqref{stop_r}.  
		We have
		\begin{align}
		\mathsf{num}(\underline{I}_i(q,q'))&\leq \alpha^i\mathsf{M}^{i+1},\label{eq:proff_compl1}\\
		\mathsf{larg}(\underline{I}_i(q,q'))&\leq g^i(p'),\label{eq:proff_compl2}
		\end{align}
		for any $q,q'\in Q_{rd} $, with $ Q_{rd}$ being defined as in~\eqref{Qrd}, where
		\begin{align}
		\alpha&: = \max_{q\in Q_{rd}}|\mathsf{out}(q)|\label{eq:alp}\\
		\mathsf{M}&:= \max_{q,q'\in Q_{rd}} \mathsf{num}(\underline{I}_0(q,q')),\label{eq:mathsfM}\\
		p'&:= \max_{\mathcal{P},\mathcal{P}\subset\underline{I}_0(q,q')\text{ with }q,q'\in Q_{rd}} \mathsf{numh}(\mathcal{P}),\label{eq:p'}
		\end{align}
		in which $|\mathsf{out}(q)|$ is the cardinality of the set 
		$$\mathsf{out}(q):=\{q'\in Q~|~\exists \sigma\in \Pi, (q,\sigma,q')\in \delta \};$$
		$\underline{I}_0$ is as in~\eqref{iter_p};
		$\mathsf{num}(\cdot)$, $\mathsf{larg}(\cdot)$, and $\mathsf{numh}(\cdot)$ are defined in~\eqref{eq:poly_num},~\eqref{eq:hyperplane_num} and~\eqref{eq:numh}, respectively;
		$\mathcal{P}$ is any arbitrary polytope within $\underline{I}_0(q,q')$;
		and $g^i:\mathbb{N}\rightarrow\mathbb{N}$, with $i\in \mathbb{N}$, is recursively defined as
		\begin{align}
		g^i(p') &= p',\text{ when }i=0;\nonumber\\
		g^i(p') &= p'+\tilde{g}_{S}(g^{i-1}(p')),\text{ when }i\geq 1,\label{def:gi}
		\end{align}
		where $\tilde{g}_{S}(\cdot)$ is defined in~\eqref{gS}.
\end{theorem}
\begin{remark}\label{rem:conser_complexity}
	As a key insight, Theorem~\ref{thm:upperbound} provides upper bounds on: 1) the number of polytopes within $\underline{I}_i(q,q')$; 2) the number of hyperplanes defining each polytope within $\underline{I}_i(q,q')$.
		These upper bounds are conservative since they are derived without considering the possibility of eliminating redundant hyperplanes and polytopes in practice.
		Concretely, intersections among polytopes in each iteration may contain some redundant hyperplanes, which can be eliminated by computing the minimal representations of these intersections~\cite{Baotic2009Polytopic}.
		Additionally, one can also reduce $\mathsf{num}(\underline{I}_i(q,q'))$ by computing unions among some of the polytopes within $\underline{I}_i(q,q')$, in case these unions are in the form of polytopes.
\end{remark}
The proof of Theorem~\ref{thm:upperbound} is provided in Appendix~\ref{proof:section5}.
Based on Theorem~\ref{thm:upperbound}, we propose the worst-case space and time complexities of Algorithm~\ref{alg} in the following corollary.
\begin{corollary}\label{thm:complexity}
	Consider a dtLCS $S=(X,X_0,U,W,f)$ with $W= \{\mathbf{0}_n\}$, a DSA $\mathcal{A} = (Q, q_0, \Pi,\delta, \text{Acc})$ modeling the desired $\omega$-regular property, and $i\in \mathbb{N}_{>0}$ the number of iterations.
	The worst-case space and time complexities of Algorithm~\ref{alg} are 
	\begin{align}
	&\mathcal{O}\Big(|\delta|\alpha^i\mathsf{M}^{i+1}g^i(p')n\Big),\label{eq:spacecomp}\\
	&\mathcal{O}\Big(\! |\delta|c_1\!\big(\alpha^{i-1}\mathsf{M}^{i},g^{i-1}(p')\big)\!+\!|\delta|\alpha^{i-1}\mathsf{M}^{i+1}c_2\big(p',\tilde{g}_{S}(g^{i-1}(p'))\big) + |\delta|c_3\big(\alpha^i\mathsf{M}^{i+1},g^i(p'),\alpha^{i-1}\mathsf{M}^{i},g^{i-1}(p')\big) \Big),\label{eq:timecomp}
	\end{align}
	respectively, in which $|\delta|$ is the number of transitions among $q,q'\in Q_{rd}$, with $Q_{rd}$ as defined in~\eqref{Qrd}; 
		$\alpha$, $\mathsf{M}$, $p'$ and $g^i(p')$ are defined in~\eqref{eq:alp}-\eqref{def:gi}, respectively;
		$\tilde{g}_{S}(\cdot)$ is defined in~\eqref{gS};
		$c_1$, $c_2$, and $c_3$ represent the computation costs for accomplishing different tasks as defined in Table~\ref{computation_cost}.
\end{corollary}
\begin{table}[]
	\caption{\label{computation_cost} Definition of $c_1$, $c_2$, and $c_3$.}
	\begin{small}
			\begin{tabular}{|l|l|}
				\hline
				Functions               & Tasks                                                                                                                                                                                                                              \\ \hline
				$c_1(a_1,b_1)$         & \begin{tabular}[c]{@{}l@{}}Compute $pre(X')$, with $X'$ being a P-collection in $\mathbb{R}^n$ for which \\ $\mathsf{num}(X')=a_1$, $\mathsf{larg}(X')=b_1$\end{tabular}                                                                                             \\ \hline
				$c_2(a_2,b_2)$         & Concatenate matrices $\mathsf{P}_1\in\mathbb{R}^{a_2\times (n+1)}$ with $\mathsf{P}_2\in \mathbb{R}^{b_2\times (n+1)}$                                                                                                             \\ \hline
				$c_3(a_3,b_3,a'_3,b'_3)$ & \begin{tabular}[c]{@{}l@{}}Check whether $X'_{i-1}\subseteq X'_{i}$ holds, with  $X'_i, X'_{i-1}\subset \mathbb{R}^n$ being P-collections, \\ where  $\mathsf{num}(X'_i)=a_3$, $\mathsf{larg}(X'_i)=b_3$, $\mathsf{num}(X'_{i-1})=a'_3$, $\mathsf{larg}(X'_{i-1})=b'_3$\end{tabular} \\ \hline
			\end{tabular}
	\end{small}	\vspace{0.5cm}
\end{table}
\begin{remark}
	For each $i\!\in\!\mathbb{N}_{>0}$, the tasks for the iteration in~\eqref{iter_p} and~\eqref{stop_r} include: 1) computing the one-step-backward projection $\textbf{P}(\underline{I}_{i-1})$ of $\underline{I}_{i-1}$; 2) computing the intersection $\underline{I}_0\cap \textbf{P}(\underline{I}_{i-1})$; 3) checking whether $\underline{I}_{i-1}\subseteq \underline{I}_{i}$ holds\footnote{One can verify that $\underline{I}_{i}\subseteq \underline{I}_{i-1}$ always holds based on the way of computing $\underline{I}_{i}$.}.
		Their computation costs correspond to the first, second, and third term in~\eqref{eq:timecomp}, respectively.
		Here, the closed-form expressions of $c_1$, $c_2$, and $c_3$ depend on the concrete methods that are deployed for their associated tasks.
		For instance, given a polytope $X'\!\subset\! \mathbb{R}^n$, computing $pre(X')$ includes the computation of inverse image of a polytope and polyhedral projection~\cite{Rakovic2006Reachability}.	
		For linear systems as in~\eqref{eq:linear_subsys}, 	the inverse image of a polytope can be obtained via simple matrix multiplications as in~\cite[Section 4]{Rungger2017Computing}, while different approaches can be used to compute the projection of a polytope~\cite{Jones2004Equality,Karavelas2016Maximum,Yu2019efficient}.
		Similarly, various results can be applied to check whether $\underline{I}_{i-1}\!\subseteq\! \underline{I}_{i}$ holds, e.g. \cite{Bemporad2001Convexity,Baotic2009Polytopic}.	 
\end{remark}
\begin{remark}
	With slight modifications, Definition~\ref{helpdef}, Theorem~\ref{thm:upperbound}, and Corollary~\ref{thm:complexity} can also be leveraged to analyze the space and time complexities of the computation of ($\varepsilon_x$,$\varepsilon_u$)-contraction-based approximation.
	Concretely,  $pre(\cdot)$ in~\eqref{gS} should be defined as in~\eqref{eq:pre_cont} (instead of~\eqref{eq:pre_orig}),
	and $\underline{I}_0$ in~\eqref{eq:mathsfM} and~\eqref{eq:p'} should be defined as in~\eqref{iter_po} (instead of~\eqref{iter_p}).
\end{remark}
The proof of Corollary~\ref{thm:complexity} is provided in Appendix~\ref{proof:section5}.
As for the closed-form expressions of $\tilde{g}_S(p)$ in~\eqref{gS} and $g^i(p')$ in~\eqref{eq:spacecomp} and~\eqref{eq:timecomp}, we have the following results.
\begin{proposition}\label{prop:minkov}
	Consider a dtLCS $S=(X,X_0,U,W,f)$ as in Definition~\ref{def:dtLCS}, where $n$ is the dimension of $X$, $W= \{\mathbf{0}_n\}$, and $p_{\mathsf{U}}:= \mathsf{numh}(BU)$.
	Given $p'$ as in~\eqref{eq:p'}, and $i\in\mathbb{N}$ the number of iterations, one has
	\begin{align}
	\tilde{g}_S(p')\leq\left\{
	\begin{aligned} 
	&\quad\quad\quad\quad 2, \quad\quad\quad\quad\quad\ \, \text{ when } n=1;\\
	&\quad\quad\quad p_{\mathsf{U}}+p',\quad\quad\quad\ \ \,\text{ when } n=2;\\
	&(4p_{\mathsf{U}}-9)p'+26-9p_{\mathsf{U}}, \,\text{ when } n=3.
	\end{aligned}\right.
	\end{align}
	Accordingly, one gets 
	\begin{align}\label{upperb_gi}
	g^i(p')\leq\left\{
	\begin{aligned} 
	&\quad\quad 2(i+1),\quad\quad\quad\ \ \text{ when } n=1;\\
	&\quad\quad p' + i(p'+p_{\mathsf{U}}),\ \ \, \text{ when } n=2;\\
	& \frac{1-\tilde{a}^{i+1}}{1-\tilde{a}}p'+\frac{1-\tilde{a}^i}{1-\tilde{a}}\tilde{b},\text{ when } n=3.
	\end{aligned}\right.
	\end{align}
	with $\tilde{a}=4p_{\mathsf{U}}-9$, and $\tilde{b}=26-9p_{\mathsf{U}}$.
\end{proposition}
Note that we have $p_{\mathsf{U}}\leq \mathsf{numh}(U)+2(n-rank(B))$ according to~\cite[Corollary 3.5]{Kerrigan2001Robust}.
Considering~\eqref{eq:pre_msum}, solving the closed-form expressions of $\tilde{g}_S(p')$ is equivalent to answering the following question: \textit{given polytopes $\mathcal{P}_1$ and $\mathcal{P}_2$ defined by $p'$ and $p_{\mathsf{U}}$ hyperplanes, respectively, what is the upper bound of the number of hyperplanes defining $\mathcal{P}_1+\mathcal{P}_2$?}
Trivially, $2$ is the upper bound for the case $n=1$.
Additionally, one has $p_{\mathsf{U}}+p'$ being the upper bound for the case $n=2$ according to~\cite[Theorem 13.5]{VanKreveld2008Computational}, and $(4p_{\mathsf{U}}-9)p'+26-9p_{\mathsf{U}}$ being the upper bound for the case $n=3$ according to~\cite[Theorem 5.2.1]{Weibel2007Minkowski}.
Then,~\eqref{upperb_gi} can accordingly be derived.
As for the cases $n\geq 4$, to the best of our knowledge, there is no result providing the upper bounds of the number of hyperplanes defining $\mathcal{P}_1+\mathcal{P}_2$ based on $p'$ and $p_{\mathsf{U}}$.
However, once the results for these upper bounds are available, the space complexities for the cases $n\geq 4$ can readily be derived based on Corollary~\ref{thm:complexity}.

Finally, we also want to point out the difficulties in having a fair comparison between those discretization-based approaches and ours in terms of worst-case space complexity.
It is well-known that the space complexities of discretization-based approaches grow exponentially with respect to the dimension of the state (and input) sets (see~\cite[Section 5-A]{Li2022Specification} for detailed discussion) since they require the discretization of the original state and input sets in order to construct the finite state and input sets.
On the one hand, the space complexity of our approaches does not have exponential growth regarding the dimensions since we do not require such discretization.
On the other hand, the complexity of our approaches grows exponentially with respect to the number of iterations in the worst case.
It is worth noting that, however, we do not observe such exponential growth in the case studies (see Figure~\ref{fig:evolution_comp}).
As a key insight, at each iteration step $i\in\mathbb{N}$, for all $q,q'\in Q_{rd}$, one can reduce $\mathsf{num}(\underline{I}_i(q,q'))$ and $\mathsf{larg}(\underline{I}_i(q,q'))$ in~\eqref{eq:proff_compl1} and~\eqref{eq:proff_compl2} by computing the minimal representations~\cite{Baotic2009Polytopic} and the union of (some of) the polytopes in $\underline{I}_i(q,q')$. 

\section{Case Study}\label{sec5}
To show the effectiveness of our results, we first simulate the running example with the HCI-based controllers, which have already been computed in Section~\ref{sec4}.
Then, we apply our results to a cruise control example.
Finally, we compare our approaches with some currently existing tools in terms of computational time.
The synthesis and simulation are performed on a computer equipped with Quad-Core Intel Core i7 (2.7 GHz) and 16 GB of RAM running macOS Big Sur (Version 11.5.2),
using \texttt{MATLAB2019b} along with multi-parametric toolbox \texttt{MPT}~\cite{Herceg2013Multia} and optimization software~\texttt{MOSEK} (version 9.3.6)~\cite{ApS2019MOSEK}.
It is also worth noting that controllers in both cases can be applied over an infinite time horizon.
The numbers of time steps for the simulation are selected only for demonstration purposes.

\subsection{Running Example}
\begin{figure}[htbp]
	\centering
	\subfigure{
		\includegraphics[width=0.55\textwidth]{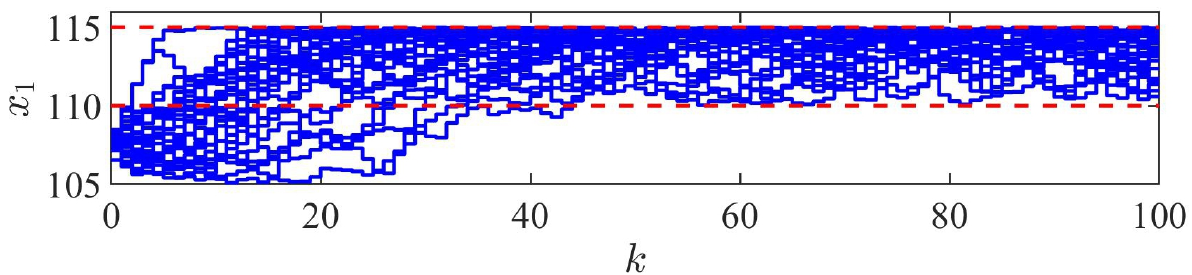}
	}
	\quad
	\subfigure{
		\includegraphics[width=0.55\textwidth]{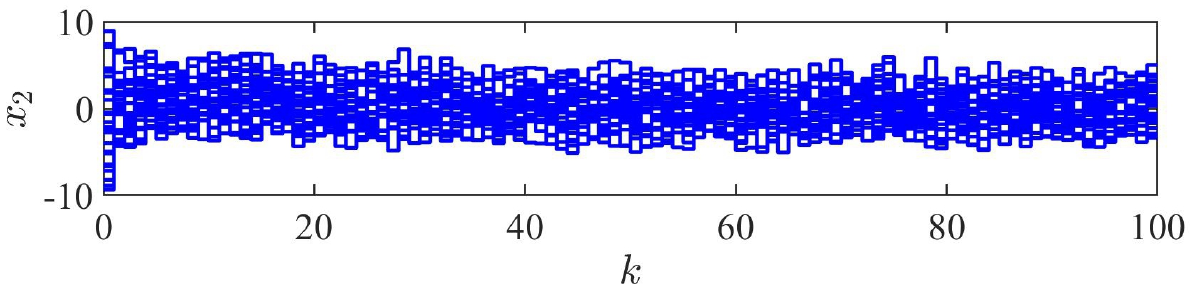}
	}
	\caption{Simulation of the running example with the controllers associated with the maximal HCI set, the ($\varepsilon_x$,$\varepsilon_u$)-contraction-based approximation and the $\varepsilon$-expansion-based approximation. }
	\label{fig:sim_cbased}
\end{figure}

Here, we randomly select 10 different initial states from $\underline{I}^*(q_0,q_0)$, $\underline{I}(\varepsilon_x,\varepsilon_u)(q_0,q_0)$, and $\underline{I}(\varepsilon)(q_0,q_0)$ (cf. Figure~\ref{fig:result_cbased} and Figure~\ref{fig:result_ebased}), respectively, and simulate the running example for 30 time steps.
In the simulation, the disturbances affecting the system are randomly generated at each time instant following a uniform distribution within the disturbance set.
The simulation results for the maximal HCI set, the ($\varepsilon_x$,$\varepsilon_u$)-contraction-based and $\varepsilon$-expansion-based approximation are shown in Figure~\ref{fig:sim_cbased}. 
One can verify that the desired property is respected.

\subsection{Cruise Control}\label{sec:61}
\begin{figure}[htbp]
	\centering
	\includegraphics[width=0.4\textwidth]{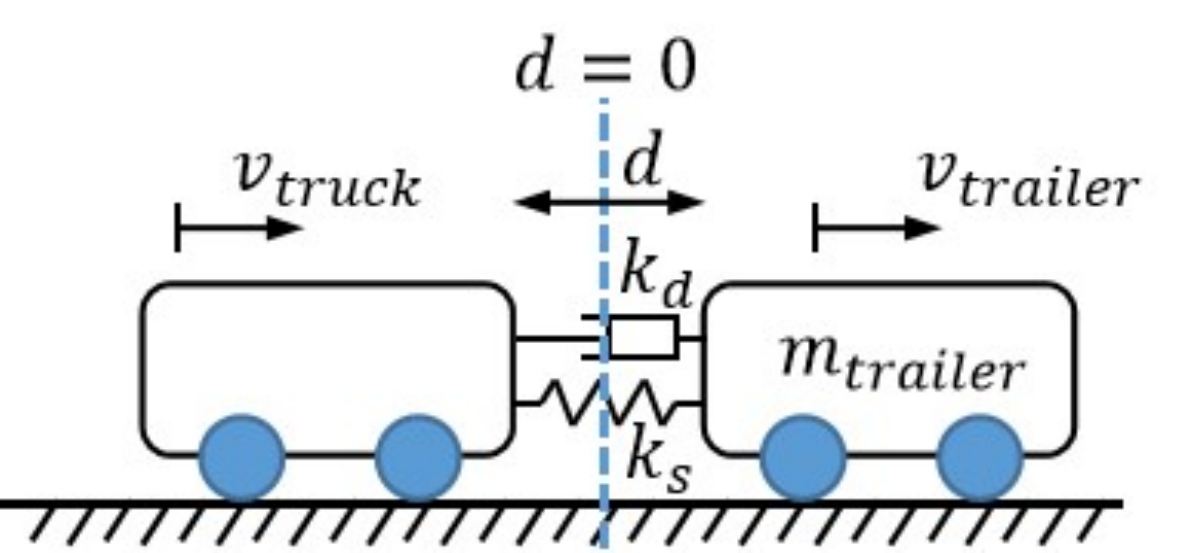}
	\caption{Cruise control problem for a truck with a trailer, with $m_{trailer}=4000$kg the mass of the trailer, $k_s=4500$N/kg and $k_d=4600$Ns/m the constants for the spring-damper system, and $d$ the distance between the truck and the trailer, where $d=0$m is the position at which there is no deformation on the spring.}
	\label{fig:truck_trailer}
\end{figure}

\begin{figure}[htbp]
	\centering
	\subfigure{
		\includegraphics[width=0.55\textwidth]{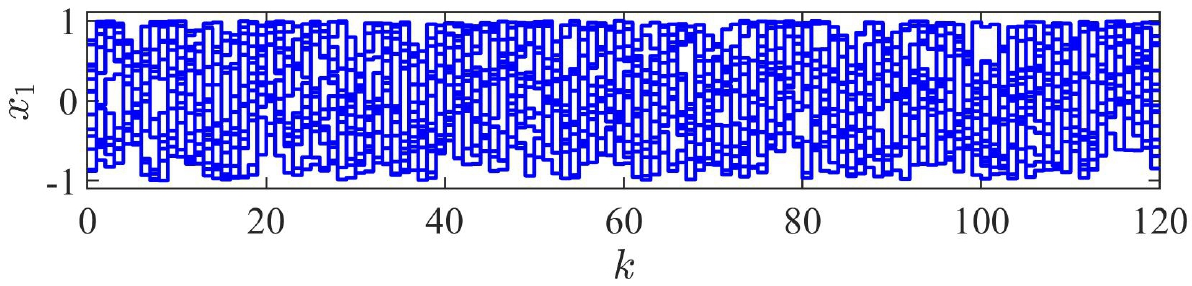}
	}
	\quad
	\subfigure{
		\includegraphics[width=0.55\textwidth]{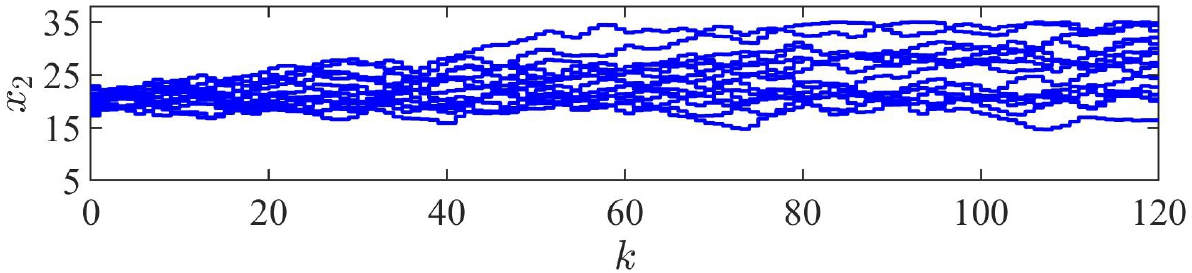}
	}
	\quad
	\subfigure{
		\includegraphics[width=0.55\textwidth]{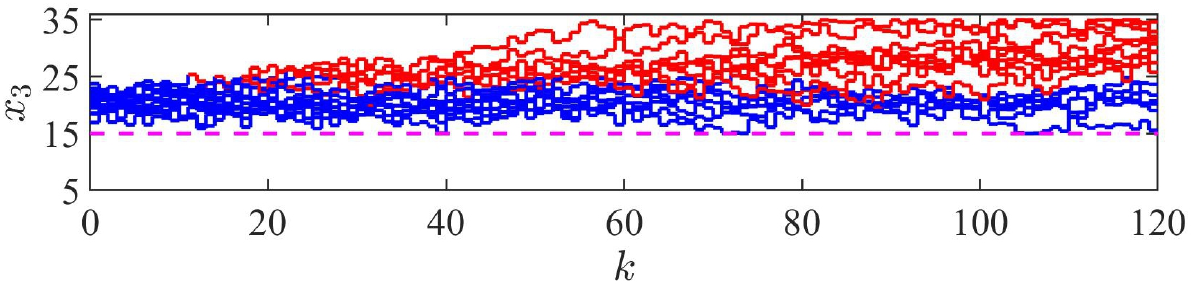}
	}
	\caption{Simulation of the cruise control problem} 
	\label{fig:sim_truck}
\end{figure}

\begin{figure}[htbp]
	\centering
	\includegraphics[width=0.55\textwidth]{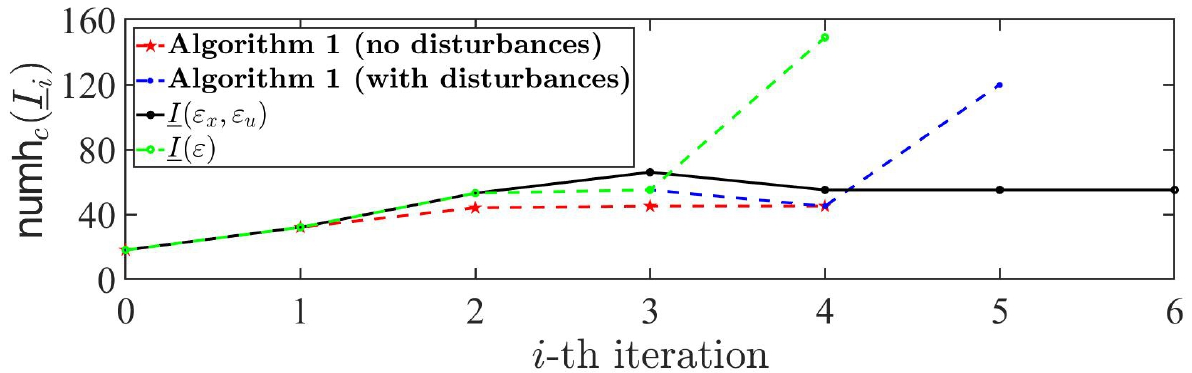}
	\caption{Evolution of the number of hyperplanes required to characterize $\underline{I}_i$, denoted by $\mathsf{numh}_{c}(\underline{I}_i)$, as $i$ increases.
	}
	\label{fig:evolution_comp}
\end{figure}

Here, we focus on a cruise control problem for a truck with a trailer as in Figure~\ref{fig:truck_trailer}, with dynamics as in~\eqref{eq:linear_subsys}, where
	\begin{align}
	A :=\begin{bmatrix}\begin{smallmatrix}0.8855\,&-0.3628& 0.3628\\0.4081\,&0.4683&0.5317\\0\,&0\,&1.0000\end{smallmatrix}\end{bmatrix}\!,\ 
	B :=\begin{bmatrix}\begin{smallmatrix}0.1018\\0.1372\\0.5000\end{smallmatrix}\end{bmatrix}\!,\label{casestudy}
	\end{align}
$x(k) = [x_1(k);x_2(k);x_3(k)]$ is the state of the system, in which $x_1(k)$, $x_2(k)$, and $x_3(k)$ are the distance between the truck and the trailer, the velocity of the trailer, and the velocity of the truck, respectively.
	Moreover, $u(k)\in[-5,5]$m/s$^2$ denotes the acceleration of the truck that is used as the control input;
	and $w(k)\in[-0.04,0.04]\times [-0.02,0.02]^2$ denotes the exogenous disturbances encompassing the model uncertainty and unexpected interferences.
	The model as in~\eqref{casestudy} is adapted from~\cite{Rungger2013Specification} by discretizing it with a sampling time $\Delta t = 0.5s$ and including exogenous disturbances. 
\begin{figure}[htbp]
	\centering
	\includegraphics[width=0.5\textwidth]{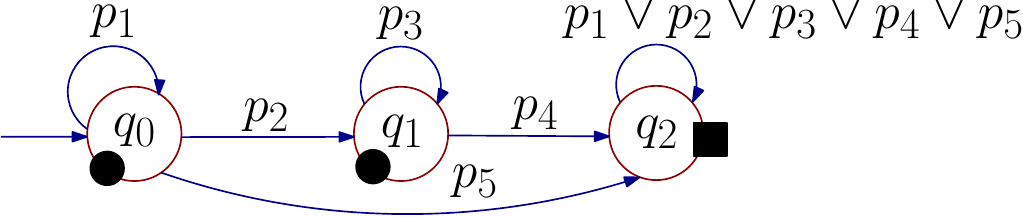}
	\caption{DSA $\mathcal{A}_q$ modeling $\psi_q$, with alphabet $\Pi=\{p_1,p_2,p_3,$ $p_4,p_5\}$; labeling function $L:X \rightarrow \Pi$ with 
			$L(x)=p_1$ when $x \in $ $ [-1,1]\times[5,35]\times[15,25]$, 
			$L(x)=p_2$ when $x \in [-1,1]\times$ $[5,35]\times(25,35]$, 
			$L(x)=p_3$ when $x \in $ $[-1,1]\times[5,35]\times[5,35]$,
			$L(x)=p_4$ when $x \in \mathbb{R}^3\backslash L^{-1}(p_3)$,
			and $L(x)=p_5$ when $x \in \mathbb{R}^3\backslash (L^{-1}(p_1)\cup L^{-1}(p_2)) $; and accepting condition $\text{Acc}=\{\langle E_1,F_1\rangle\}$, with $E_1= \{ q_3\}$, $F_1 = \emptyset$. 
			The temporal logics formula for $\psi_q$ is given by $G((p_1 U p_2)\wedge (\neg p3))$.	 
	}
	\label{fig:quadrotor_dsa}
\end{figure}
In this case study, the distance between the truck and the trailer should be within $[-1,1]$m to protect the spring-damper system, and the velocity of the truck and the trailer should be within $[5,35]$m/s due to the traffic rules.
	Additionally, to increase the throughput of the road traffic, the truck is not allowed to move slower than $15$m/s unless it has moved faster than $25$m/s.
	Such a property, denoted by $\psi_q$, can be modeled by a DSA $\mathcal{A}_q$ as depicted in Figure~\ref{fig:quadrotor_dsa}.
	To synthesize controllers enforcing $\psi_q$, we select $\underline{E}:= \cup_{\forall q'\in Q}\big(q',q_2,\underline{X}(q',q_2)\big)$.
	Additionally, to ensure the compactness of $\underline{X}\backslash\underline{E}$, we slightly deflate $\underline{X}\backslash\underline{E}$ such that $\underline{X}\backslash\underline{E}(q_0,q_1):=[-1,1]\times[5,35]\times[20+\epsilon,35]$, with $\epsilon=0.001$.
	The results of controller synthesis are summarized in Table~\ref{sim_results}.
	Then, we randomly select 10 initial states from $\underline{I}^*(q_0,q_0)$, $\underline{I}(\varepsilon_x,\varepsilon_u)(q_0,q_0)$, and $\underline{I}(\varepsilon)(q_0,q_0)$, respectively, and simulate the systems for 60 seconds (i.e. 120 time steps).
	Moreover, the disturbances are randomly generated at each time step following a uniform distribution within the disturbance set.
	The simulation results are shown in Figure~\ref{fig:sim_truck}, indicating that the desired property is enforced (note that trajectories of $x_3$ become red after $x_3$ has been larger than $25$m/s).
	Additionally, Figure~\ref{fig:evolution_comp} shows that there is no exponential growth as in~\eqref{eq:proff_compl1} and~\eqref{eq:proff_compl2} in this case study.
\begin{table}[]
	\centering
	\caption{\label{sim_results} Synthesizing controllers for the cruise control problem by computing: 1) maximal HCI set $\underline{I}^*$; 2) contraction-based approximation $\underline{I}(\varepsilon_x,\!\varepsilon_u)$ with $n \!= \!3$ and $\varepsilon_x=\varepsilon_u\!=\!0.036$; 3) expansion-based approximation $\underline{I}(\varepsilon)$ with $\varepsilon = 0.002$.}
	\renewcommand\arraystretch{1.2}
	\begin{small}
			\begin{tabular}{|c|c|c|c|}
				\hline
				& $\underline{I}^*$ & $\underline{I}(\varepsilon_x,\varepsilon_u)$              & $\underline{I}(\varepsilon)$                                                              \\ \hline
				Number of iterations                                                & 5        & 6                     & 4                                                                     \\ \hline
				Computation time (s)                                                & 21.34    	&  19.59                & 16.12                                                                 \\ \hline
				Number of hyperplanes                                               & 120      & 259                   & 149                                                                   \\ \hline
			\end{tabular}
	\end{small}\vspace{0.2cm}
\end{table}
\noindent

\subsection{Comparison with Existing Results}\label{sec:compare}
\begin{table*}[]
	\centering
	\caption{\label{tbl:compare2} Comparison among the proposed HCI-based methods and existing results in terms of computation time for synthesizing controllers enforcing: 1) (Case 1) $\psi_q$ over system  in~\eqref{casestudy} with disturbances; 2) (Case 2) $\psi_q$ over system in~\eqref{casestudy} without disturbances; 3) (Case 3) system in~\eqref{casestudy} with disturbances reaching a target set within 3 time steps.}
	\renewcommand\arraystretch{1.2}
	\begin{small}
		\begin{tabular}{|c|c|c|c|c|c|c|c|c|}
				\hline
				Methods                         & \begin{tabular}[c]{@{}c@{}}Maximal \\ HCI-set \end{tabular} & ($\varepsilon_x$,$\varepsilon_u$)-contraction  & $\varepsilon$-expansion & \texttt{ROCS} & \texttt{TuLiP} & \texttt{OmegaThreads} & CBF-based & HJ-based \\ \hline
				Case 1  & 21.34 s   & 19.59 s    	& 16.12 s   & N/A   	&$>$6 h  	& 2899.85 s  & N/A	& N/A      \\ \hline
				Case 2 	& 2.92 s  	& 7.44 s 		& 7.04 s  	& $>$6 h    &$>$6 h 	& 1933.60 s  & N/A  & N/A      \\ \hline
				Case 3  & 51.02 s 	& 86.27 s 		& 44.30 s 	& N/A  		&$>$6 h		& 7123.81 s  & N/A	& 415.15 s\\ \hline
		\end{tabular}
	\end{small}\vspace{0.5cm}
\end{table*}
In this subsection, we compare the proposed set-based approaches with existing results in terms of computation time for synthesizing controllers, including symbolic techniques (\texttt{OmegaThreads}~\cite{Khaled2021OmegaThreads} and \texttt{TuLiP}~\cite{Filippidis2016Control}), interval-analysis-based approaches (\texttt{ROCS}~\cite{Li2018ROCS}), CBF-based approaches~\cite{Jahanshahi2021Compositional}, and HJ-based approaches (\texttt{helperOC}~\cite{Bansal2017Hamilton} equipped with \texttt{toolboxLS}~\cite{Mitchell2005toolbox}).
	Moreover, since interval-based approaches do not handle systems with exogenous disturbances~\cite[Section 2.D]{Li2022Specification}, and HJ-based approaches do not handle $\omega$-regular properties, for a fair comparison among these approaches, we consider three different cases: 
	1) enforcing $\psi_q$ in Session~\ref{sec:61} over the system in~\eqref{casestudy}; 
	2) enforcing $\psi_q$ over the system in~\eqref{casestudy}, but \emph{without exogenous disturbance}; 
	3) ensuring the system in~\eqref{casestudy} reaches the region $[-1,1]\times[5,35]\times[25,35]$ from the region $[-1,1]\times[5,35]\times[5,25]$ within 3 time steps.

Following the same settings as in Table~\ref{sim_results}, we choose $n=3$ and $\varepsilon_x\!=\!\varepsilon_u\!=\!0.036$ to compute the ($\varepsilon_x$,$\varepsilon_u$)-contraction-based approximation of the maximal HCI-set, and select $\varepsilon = 0.002$ to compute the $\varepsilon$-expansion-based approximation.
	For applying \texttt{ROCS}, we select $\varepsilon=0.001$ and $\mu=0.001$ as the lower bounds of discretization parameters for state and input sets, respectively, (see~\cite[Section 4-A]{Li2022Specification} for their definitions) for a fair comparison with the setting of $\varepsilon$-expansion-based approach~\cite[Lemma 1 and Theorem 1]{Li2022Specification}.
	Moreover, considering the limitation of our computer, $0.2$ is used as the discritization parameter for discretizing the state and input sets when deploying \texttt{OmegaThreads}, \texttt{TuLiP}, and \texttt{helperOC}.
	The computation time for synthesizing controllers with different approaches is summarized in Table~\ref{tbl:compare2}, which indicates that our approaches require less computation time than other ones.
	Concretely, $>$6 h means that the corresponding synthesis procedures did not terminate within 6h, and that the actual computation time is undecided.
	Additionally, when applying \texttt{OmegaThreads}, no controller was found in all cases with the current discretization parameters.
	Therefore, smaller discretization parameters for the state and input sets are needed to potentially obtain controllers, which would, however, result in longer computation time.
	As for using CBF-based methods in~\cite{Anand2021Compositional}, although we set the potential control barrier function, the multipliers, and the controller as polynomials of up to degree eight, no controller was found in any cases.

\section{Conclusion}\label{sec6}
In this paper, we proposed for the first time a notion of so-called hybrid controlled invariant set (HCI set), based on which we synthesize controllers to enforce $\omega$-regular properties over linear control systems affected by bounded disturbances. 
Given a linear control system and a deterministic Streett automata (DSA) modeling the desired $\omega$-regular property, we first construct a product system between the linear control system and the DSA.
Then, we compute the maximal HCI set by utilizing a set-based approach over the hybrid state set of the product system.
Additionally, we provide two approaches to compute approximations of the maximal HCI sets within a finite number of iterations: one by deflating the original state and input sets, the other by expanding the disturbance set. 
The effectiveness of our methods is shown by two case studies, and by comparison with existing tools.

\bibliographystyle{elsarticle-num}   
\bibliography{bib}

\appendix

\renewcommand{\theequation}{A.\arabic{equation}}
\section{Proof of Statements}
\subsection{Proof of Proposition~\ref{lem:stationary_policy} and Theorem~\ref{thm:convergence}}\label{proof:section3}
We first show the results for Proposition~\ref{lem:stationary_policy}.

{\bf Proof of Proposition~\ref{lem:stationary_policy}}
Consider any controller sequence $\mu' = \{\mu'_0,\mu'_1,\ldots,\mu'_i,\ldots\}$, with $i\in\mathbb{N}$, associated with $\underline{I}$ such that for any initial state $\underline{x}(0)\in \underline{I}$, and infinite state sequence $\underline{\xi} = \{\underline{x}(0),\underline{x}(1),\ldots,\underline{x}(i),\ldots\}$, one has $\underline{x}(i)\!\in\!\underline{I}, \forall i\!\in\!\mathbb{N}$, when $\mu'$ is applied.
Note that such controller sequence exists according to the definition of the HCI set as in Definition~\ref{def:HCI}.
Hence, at time instant $i=0$, $\forall \underline{x}\in\underline{I}$, $\forall w\in\underline{W}$, one gets $\underline{x}':=\underline{f}(\underline{x},u,w)\in\underline{I}$ with $u = \mu'_0(\underline{x})$.
Since one has $\underline{x}'\in\underline{I}$, then $\forall w'\in\underline{W}$, we again have $\underline{x}'':=\underline{f}(\underline{x}',u,w')\in\underline{I}$ with $u = \mu'_0(\underline{x})$ at time instant $i=1$.
Therefore, one can verify that with the sequence of controller $\mu'':=\{\mu''_0,$ $\mu''_1,\ldots,\mu''_i,\ldots \}$ with $\mu''_i=\mu'_0$, $\forall i\in\mathbb{N}$, one also has $\underline{x}(i)\!\in\!\underline{I}, \forall i\!\in\!\mathbb{N}$, when $\mu''$ is utilized.
Note that $\mu''$ is a stationary controller as in Definition~\ref{def:HCIcontroller}, which completes the proof.$\hfill\blacksquare$

Next, we proceed with showing Theorem~\ref{thm:convergence}, for which some additional definitions and lemmas are required.
First, we define a set
\begin{align}
G(\underline{X}')\! :=\! \{(\underline{x},u)\!\in\! \underline{X}\times \underline{U}~|~\forall w\in\underline{W},\underline{f}(\underline{x},u,w)\!\in\! \underline{X}'\},\label{def:Gi}
\end{align}
where $\underline{X}'\subseteq \underline{X}$.
Accordingly, consider $\underline{I}_0$ along with the iteration of $\underline{I}_i$ as in~\eqref{iter_p}, we define $G_i$ with $i\in\mathbb{N}_{>0}$ as:
\begin{align}
G_1(\underline{I}_0) &:= G(\underline{I}_0);\label{eq:G1}\\
G_i(\underline{I}_0) &:= G(\underline{I}_{i-1}),\,i\geq 2.\label{eq:Gi}
\end{align}
Now, based on theses definitions, we propose Lemma~\ref{lem:compact} and Lemma~\ref{lem:crt_eq}, which are also necessary for showing the proof of Theorem~\ref{thm:convergence}.
\begin{lemma}\label{lem:compact}
	Consider a dtLCS $S$ as in~\eqref{eq:linear_subsys}, a DSA $\mathcal{A}$ modeling the desired $\omega$-regular property, and the product dtLCS $S\otimes\mathcal{A} = (\underline{X}, \underline{X}_0,\underline{U},\underline{W},\underline{f})$ such that Assumption~\ref{assum1} holds. 
	Then, $G_i(\underline{I}_0)$ are compact for all $i\in\mathbb{N}_{>0}$, with $G_i$ as defined in~\eqref{eq:G1} and~\eqref{eq:Gi}, and $\underline{I}_0$ as in~\eqref{iter_p}.
\end{lemma}
{\bf Proof of Lemma~\ref{lem:compact}}
In case that $G_i(\underline{I}_0)=\emptyset$, the assertion of Lemma~\ref{lem:compact} holds trivially.
Therefore, we assume that $G_i(\underline{I}_0)\neq\emptyset$, for all $i\in\mathbb{N}_{>0}$.
To this end, we first show that $\underline{I}_i$ are compact for all $i\in\mathbb{N}$.
Then, we show that $G_i(\underline{I})$ is compact when $\underline{I}$ is compact, and the compactness of $(G_i(\underline{I}_0))_{i\in\mathbb{N}_{>0}}$ follows by the compactness of $(\underline{I}_i)_{i\in\mathbb{N}}$.
Firstly, $\underline{I}_0$ as in~\eqref{iter_p} is compact according to Assumption~\ref{assum1}.
Since intersection of two compact sets are still compact, we show the compactness of $(\underline{I}_i)_{i\in\mathbb{N}}$ by showing $\textbf{P}(\underline{I})$ as in~\eqref{eq:iter_pf} is compact if $\underline{I}$ is compact.
For this purpose, we rewrite $\textbf{P}(\underline{I})$ as $\textbf{P}(\underline{I}) = \{\underline{x}\in \underline{X}~|~ \exists u \in \underline{U},\text{ s.t. }\underline{f}(\underline{x},u,\mathbf{0}_n)\in \underline{I}'\}$,
with $\underline{I}'$ a bounded set being defined as $\underline{I}' = \{\underline{x}'~|~\{\underline{x}'\}\oplus \underline{W}\subseteq \underline{I}\}\label{I'}$.
Consider any $\underline{x}'':=(q,q',x')\notin \underline{I}'$.
By definition of $\underline{I}'$, there exists at least one $w\in\underline{W}$ such that one gets $\underline{z}:=(q,q',x'+w)\notin\underline{I}$.
Since $\underline{I}$ is compact (and therefore closed), then there exists an open ball $\mathsf{B}$ in the sense of the distance as defined in~\eqref{def:distance} centered at $(q,q',0)$ such that $(\underline{z}+\mathsf{B})\cap\underline{I}=\emptyset$, with $\underline{z}+\mathsf{B}$ denotes the Minkowski sum between $\underline{z}$ and $\mathsf{B}$.
Accordingly, for any $\underline{x}'':=(q,q',x')\notin \underline{I}'$, one gets $(\underline{x}''+\mathsf{B})\cap\underline{I}'=\emptyset$, which implies that $\underline{I}'$ is closed.
Therefore, $\forall q,q'\in Q$ such that $\underline{I}'(q,q')\neq \emptyset$, $\underline{I}'(q,q')$ is closed and bounded (and therefore compact).
Note that  $Q$ is a finite set, and finite union of compact sets is still compact.
Hence, it is straightforward that $\underline{I}'=\cup_{q,q'\in Q}\underline{I}'(q,q')$ is also compact.
Since the dynamics of dtLCS $S$ as in~\eqref{eq:linear_subsys} is continuous, mapping $\underline{f}$ is also continuous.
Then, the compactness of $\textbf{P}(\underline{I})$ follows by the compactness of $\underline{U}$, $\underline{I}$, and $\underline{I}'$.

As for the compactness of $G_i(\underline{I})$ given $\underline{I}$ is compact, we rewrite $G_i(\underline{I})$ as in~\eqref{def:Gi} as $G(\underline{I})\! =\! \{(\underline{x},u)\!\in\! \underline{X}\times \underline{U}~|~\underline{f}(\underline{x},u,\mathbf{0}_n)\in \underline{I}'\}$.
Then, the compactness of $G(\underline{I})$ can be proved similarly to that of the compactness of $\textbf{P}(\underline{I})$, which completes the proof.
$\hfill\blacksquare$

\begin{lemma}\label{lem:crt_eq}
	Consider $G_i$ as defined in~\eqref{eq:G1} and~\eqref{eq:Gi}, and $\underline{I}_0$ as in~\eqref{iter_p}.
	We have
	\begin{align}
	\pi_{\underline{X}}\Big(\bigcap_{i=1}^{\infty} G_i(\underline{I}_0)\Big) = \lim\limits_{i\rightarrow \infty} \underline{I}_i,\label{eq:003}
	\end{align}
	with $\pi_{\underline{X}}(G_i(\underline{I}_0))$ the projection of $G_i(\underline{I}_0)$ on to $\underline{X}$.
\end{lemma}
{\bf Proof of Lemma~\ref{lem:crt_eq}}
To show Lemma~\ref{lem:crt_eq}, we show that 1) $\pi_{\underline{X}}\Big(\bigcap_{i=1}^{\infty} G_i(\underline{I}_0)\Big) \subseteq \lim\limits_{i\rightarrow \infty} \underline{I}_i$; 2) $\lim\limits_{i\rightarrow \infty} \underline{I}_i\subseteq \pi_{\underline{X}}\Big(\bigcap_{i=1}^{\infty} G_i(\underline{I}_0)\Big)$.

First, we show that 
\begin{equation}
\pi_{\underline{X}}\Big(\bigcap_{i=1}^{\infty} G_i(\underline{I}_0)\Big) \subseteq \lim\limits_{i\rightarrow \infty} \underline{I}_i,\label{eq1}
\end{equation}
holds.
Let's denote by $\underline{\xi} = \{\underline{x}(0),\underline{x}(1),\ldots,\underline{x}(i),\ldots\}$ an infinite state sequence of $S\otimes\mathcal{A}$.
On one hand, according to the definition of $G_i(\underline{I}_0)$, $\pi_{\underline{X}}\Big(\bigcap_{i=1}^{\infty} G_i(\underline{I}_0)\Big)$ denotes the set of $\underline{x}\in\underline{I}_0$, from which there exists a stationary controller $\bar{\mu} = \{\mu,\mu,\ldots\}$ such that $\underline{x}(i)\in\underline{I}_0$, for all $i\in\mathbb{N}$.
On the other hand, according to the iteration in~\eqref{iter_p}, $\lim\limits_{i\rightarrow \infty} \underline{I}_i$ denotes the set of all $\underline{x}\in\underline{I}_0$, from which there exists a controller (either stationary or non-stationary) $\bar{\mu}' = \{\mu'_1,\mu'_2,\ldots\}$ such that $\underline{x}(i)\in\underline{I}_0$ for all $i\in\mathbb{N}$.
Therefore,~\eqref{eq1} holds.

Next, we show that 
\begin{equation}
\lim\limits_{i\rightarrow \infty} \underline{I}_i\subseteq \pi_{\underline{X}}\Big(\bigcap_{i=1}^{\infty} G_i(\underline{I}_0)\Big), \label{eq2}
\end{equation}
holds.
According to the definition of $G_i(\underline{I}_0)$ and $\underline{I}_i$, one gets 
\begin{align}
\lim\limits_{i\rightarrow \infty} \underline{I}_i=\bigcap_{i=1}^{\infty} \pi_{\underline{X}}(G_i(\underline{I}_0)).\label{eq:002}
\end{align}
Therefore, we proceed with proving 
\begin{equation}
\bigcap_{i=1}^{\infty} \pi_{\underline{X}}(G_i(\underline{I}_0))\subseteq \pi_{\underline{X}}\Big(\bigcap_{i=1}^{\infty} G_i(\underline{I}_0)\Big).\label{eq01}
\end{equation}
Consider an $\underline{x}\in \bigcap_{i=1}^{\infty} \pi_{\underline{X}}(G_i(\underline{I}_0))$.
Then, there exists a sequence $\{u_i\}_{i\in\mathbb{N}}$, such that $(\underline{x},u_i)\in G_i(\underline{I}_0)$, $\forall i\in\mathbb{N}$.
On one hand, according to the computation of $\underline{I}_i$, $i\in\mathbb{N}$ as in~\eqref{iter_p}, it is straightforward that $\underline{I}_0\supseteq \underline{I}_1 \supseteq \ldots \supseteq \underline{I}_i \supseteq \ldots$.
Then, considering the definition of $G_i((\underline{I}_0))$ as in~\eqref{eq:G1} and~\eqref{eq:Gi}, for all $i\in\mathbb{N}_{>0}$, one has $G_1(\underline{I}_0) \supseteq G_2(\underline{I}_0) \supseteq \ldots \supseteq G_i(\underline{I}_0) \supseteq \ldots$.
Hence, $\forall i'\geq i>0$, if one has $(\underline{x},u_{i'})\in G_{i'}(\underline{I}_0)$, then one gets $(\underline{x},u_{i'})\in G_{i}(\underline{I}_0)$.
On the other hand, since $G_i(\underline{I}_0)$ are compact according to Lemma~\ref{lem:compact}, any sequences of elements within $G_i(\underline{I})$ has at least one limit point $(\underline{x},u)\in G_i(\underline{I})$.
This indicate that $\exists (\underline{x},u)\in G_i(\underline{I}_0)$, $\forall i\in\mathbb{N}_{>0}$, i.e., one has $(\underline{x},u)\in\bigcap_{i=1}^{\infty} G_i(\underline{I}_0)$.
This indicates that $\underline{x}\in \pi_{\underline{X}}\Big(\bigcap_{i=1}^{\infty} G_i(\underline{I}_0)\Big)$, which implies that~\eqref{eq01} holds, and as a result~\eqref{eq2} holds.
Then, we complete the proof by combining~\eqref{eq1} and~\eqref{eq2}. $\hfill\blacksquare$

With Lemma~\ref{lem:compact}, Lemma~\ref{lem:crt_eq}, and Proposition~\ref{lem:stationary_policy}, we are ready to prove Theorem~\ref{thm:convergence}.

{\bf Proof of Theorem~\ref{thm:convergence}}
In case that $\underline{I}^*=\emptyset$, then there exists $i\in\mathbb{N}$ such that for all $i'\geq i$, $\underline{I}_{i'}=\emptyset$ according the iteration as in~\eqref{iter_p}. 
Therefore, $\underline{I}^* = \lim\limits_{i\rightarrow \infty} \underline{I}_i$ holds trivially.
This assertion can be proved by contradiction.
Suppose $\underline{I}':=\lim\limits_{i\rightarrow \infty} \underline{I}_i\neq \emptyset$.
Then, $\forall \underline{x}\in\underline{I}'$, there exists an infinite sequence of inputs $\xi_u(u(0),u(1),\ldots)$ such that the corresponding infinite state sequence $\xi_x(x(0),x(1),\ldots)$ can be enforced within $\underline{I}_0$, i.e., $\underline{I}'$ is an HCI set for $S\otimes\mathcal{A}$.
However, this is contradictory to the fact that the maximal HCI set $\underline{I}^*$ is empty.

Next, we consider the case in which $\underline{I}^*\neq\emptyset$.
Considering~\eqref{eq:002}, we first show that $\bigcap_{i=1}^{\infty} \pi_{\underline{X}}(G_i(\underline{I}_0))$ is an HCI set for $S\otimes\mathcal{A}$, which implies that
\begin{equation}
\bigcap_{i=1}^{\infty} \pi_{\underline{X}}(G_i(\underline{I}_0))\subseteq\underline{I}^*,\label{eq3}
\end{equation}
holds.
Consider a controller $\mu:\underline{X}\rightarrow \underline{U}$ such that for all $\underline{x}\in\bigcap_{i=1}^{\infty} \pi_{\underline{X}}(G_i(\underline{I}_0))$, $(\underline{x},\mu(\underline{x}))\in\bigcap_{i=1}^{\infty} G_i(\underline{I}_0)$ (such controller exists according to Lemma~\ref{lem:crt_eq} by considering~\eqref{eq:003} and~\eqref{eq:002}).
Then, by definition of $G_i(\underline{I}_0)$, $\forall \underline{x}\in\bigcap_{i=1}^{\infty} \pi_{\underline{X}}(G_i(\underline{I}_0))$, and $\forall w\in\underline{W}$, one gets
$\underline{f}(\underline{x},\mu(\underline{x}),w)\in \bigcap_{i=1}^{\infty} \pi_{\underline{X}}(G_i(\underline{I}_0))$.
Therefore, $\bigcap_{i=1}^{\infty} \pi_{\underline{X}}(G_i(\underline{I}_0))$ is an HCI set for $S\otimes\mathcal{A}$ so that~\eqref{eq3} holds according to Definition~\ref{def:HCI}.

Next, we proceed with showing that
\begin{equation}
\underline{I}^*\subseteq\bigcap_{i=1}^{\infty} \pi_{\underline{X}}(G_i(\underline{I}_0)),\label{eq4}
\end{equation}
also holds.
On one hand, according to Proposition~\ref{lem:stationary_policy}, there exists a HCI-based controller $\mu$, such that for all $\underline{x}\in\underline{I}^*$, and for all $w\in\underline{W}$, one gets $\underline{f}(\underline{x},\mu(\underline{x}),w)\in\underline{I}^*$.
On the other hand, by definition of $G_i(\underline{I}_0)$ and the HCI-based controller, one has $(\underline{x},\mu(\underline{x}))\in\bigcap_{i=1}^{\infty} G_i(\underline{I}_0)$, indicating that $\underline{x}\in\pi_{\underline{X}}\Big(\bigcap_{i=1}^{\infty} G_i(\underline{I}_0)\Big)$.
Meanwhile, by~\eqref{eq:003} and~\eqref{eq:002}, one has $\underline{x}\in \bigcap_{i=1}^{\infty} \pi_{\underline{X}}(G_i(\underline{I}_0))$, and therefore~\eqref{eq4} also holds.
Then, we are able to complete the proof by combining~\eqref{eq:002},~\eqref{eq3}, and~\eqref{eq4}.
$\hfill\blacksquare$

\subsection{Proof of Lemma~\ref{lem:cxcu} and Corollary~\ref{col:cxcu}}\label{proof:section4.1}
First, we propose Proposition~\ref{prop:help4.3} that facilitates the proof of Lemma~\ref{lem:cxcu} and Corollary~\ref{col:cxcu}.
\begin{proposition}\label{prop:help4.3}
	If $\exists c_x,c_u\in\mathbb{R}_{>0}$ and $n'\in\mathbb{N}$ such that for all $x\in\mathbb{R}^n$, there exists $\nu:[0,n']\rightarrow\mathbb{R}^m$ with which the following conditions hold:
	\begin{itemize}
		\item (\textbf{Cd.1}) $\xi_{x}(0)=x$ and $\xi_{x}(n')=\mathbf{0}_n$ with $\xi_{x}(k+1)=A\xi_{x}(k)+B\nu(k)$ for all $k\in[0,n']$;
		\item (\textbf{Cd.2}) $\nu(k)\leq c_u|x|$ holds for all $k\in[0,n']$;
		\item (\textbf{Cd.3}) $\xi_{x}(k)\leq c_x|x|$ holds for all $k\in[0,n']$;
	\end{itemize} 
	then, for all $\gamma\in\mathbb{R}_{>0}$, one has $\gamma\mathbb{B}^n\subseteq \mathcal{N}_{n'}(\varepsilon_x,\varepsilon_u)$, with $\varepsilon_x=c_x\gamma$ and $\varepsilon_u=c_u\gamma$.
\end{proposition}
{\bf Proof of Proposition~\ref{prop:help4.3}}
According to Definition~\ref{def:seq:Ni}, (Cd.1) in Proposition~\ref{prop:help4.3} indicates that there exists some $\varepsilon'_x,\varepsilon'_u\in\mathbb{R}_{>0}$ such that $\xi_{x}(t)\in\mathcal{N}_{n'-t}(\varepsilon'_x,\varepsilon'_u)$, and then (Cd.2) as well as (Cd.3) guarantee that  $\xi_{x}(t)\in\mathcal{N}_{n'-t}(\varepsilon_x,\varepsilon_u)$ with $\varepsilon_x = c_x|x|$ and $\varepsilon_u = c_u|u|$.
Therefore, one has $x\in|x|\mathbb{B}^n\subseteq \mathcal{N}_{n'}(\varepsilon_x,\varepsilon_u)$, which completes the proof.$\hfill\blacksquare$	

Now, we are ready to show the proof of Lemma~\ref{lem:cxcu}.

{\bf Proof of Lemma~\ref{lem:cxcu}}
The proof of Lemma~\ref{lem:cxcu} is given by leveraging Proposition~\ref{prop:help4.3}.
Concretely, we show the existence of $c_x$ and $c_u$ when $n'=n$ such that (Cd.1), (Cd.2) and (Cd.3) are fulfilled.
Considering any $x\in\mathbb{R}^n$, (Cd.1) requires that there exists a control sequence 
\begin{equation}
\nu = [\nu(n-1)^T;\nu(n-2)^T;\ldots;\nu(0)^T],\label{eq:ctrsq}
\end{equation}
with $\nu(k)\in\mathbb{R}^m$ for all $k\in[0,n-1]$, such that $\xi_x(n) = A^nx+\mathcal{C}\nu=\mathbf{0}_n$, with $\mathcal{C}=[B;AB;\ldots;A^{n-1}B]^T$ the controllability matrix.
Since $(A,B)$ is controllable, one has $rank(\mathcal{C})=n$, indicating the existence of such control sequence.
Therefore, (Cd.1) holds.
Let $\mathcal{C}'\in\mathbb{R}^{n\times n}$ be a matrix that contains $n$ linearly independent columns of $\mathcal{C}$.
Here, we select $\nu$ as in~\eqref{eq:ctrsq} by setting the entries $\underline{\nu}$ of $\nu$ associated with $\mathcal{C}'$ of $\mathcal{C}$ as $\underline{\nu}=-(\mathcal{C}')^{-1}A^{n}x$, and the remaining entries of $\nu$ as zero.
Accordingly, one can verify that $\xi_x(n) = A^nx+\mathcal{C}\nu=\mathbf{0}_n$ holds with such $\nu$.
In this case, since $|\underline{\nu}|\leq|(\mathcal{C}')^{-1}A^n||x|$ holds, (Cd.2) also holds with 
\begin{equation}
c_u=|(\mathcal{C}')^{-1}A^n|.\label{eq:cuselect}
\end{equation}
Meanwhile, by applying the same $\nu$, one obtains 
$\xi_x(k) = A^kx+\sum_{t'=0}^{k-1}A^{k-t'-1}B\nu(t')$.
Accordingly, one has
\begin{small}
	\begin{align*}
	|\xi_x(k)| &= |A^kx+\sum_{t'=0}^{k-1}A^{k-t'-1}B\nu(t')|\\
	&\leq |A^k||x|+|\sum_{t'=0}^{k-1}A^{k-t'-1}B||\nu(t')|\\
	&\leq (|A^k|+|\sum_{t'=0}^{k-1}A^{k-t'-1}B|c_u)|x|,
	\end{align*}
\end{small}
with $c_u$ as in~\eqref{eq:cuselect} and $|A^k|$ the infinity norm of matrix $A^k$.
Hence, (Cd.3) holds with $c_x=\max_{k\in[0,n]}$ $(|A^k|+|\sum_{t'=0}^{k-1}A^{k-t'-1}B|c_u)$, which completes the proof.$\hfill\blacksquare$

{\bf Proof of Corollary~\ref{col:cxcu}}
Consider $c_x$, $c_u$, $n'$, and $u_j$ with $j\in[0,n'-1]$ such that~\eqref{eq:cd1} to~\eqref{eq:cd3} holds.
We prove Corollary~\ref{col:cxcu} by showing that (Cd.1), (Cd.2) and (Cd.3) in Proposition~\ref{prop:help4.3} also hold for all $x\in\mathbb{R}^n$ with the same $c_x$, $c_u$, and $n'$.
For any $x'\in\mathbb{R}^n$ with $|x'| = \beta$ and $\beta \in\mathbb{R}_{\geq0}$, we consider $u'_j\leq \beta u_j$ with $|u_j|\leq c_u$ for all $j\in[0,n'-1]$, and $z'_i\in\mathbb{R}^n$, with $i\in[1,2^n]$, which are the vertices of the $\beta\mathbb{B}^n$.
Firstly, one has
\begin{equation}
A^{n'}\!\!z'_i+\!\sum_{j=0}^{n'-1}\!A^{n'-j-1}Bu'_j=\beta (A^{n'}\!\!z_i+\!\sum_{j=0}^{n'-1}\!A^{n'-j-1}Bu_j) = \mathbf{0}_n.\label{1}
\end{equation}
As a result, (Cd.1) holds for all $z'_i$, with $i\in[1,2^n]$.
Secondly, one also has
\begin{equation}
|u'_j|=|\beta u_j|\leq c_u \beta,\forall j\in[0,n'-1],\label{1.5}
\end{equation}
which implies that condition (Cd.2) holds.
Finally, for all $d\in[1,n'-1]$, one can verify that
\begin{align}
|A^dz'_i\!+\!\sum_{j=0}^{d-1}A^{d-j-1}Bu'_j|\!\leq\! |A^dz_i\!+\!\sum_{j=0}^{d-1}\!A^{d-j-1}Bu_j||\beta|\!\leq\! c_x \beta,\label{2}
\end{align}
hold.
Hence, (Cd.3) also holds for all $z'_i$, with $i\in[1,2^n]$.
Note that due to the convexity of $\beta\mathbb{B}^n$ and the linearity of~\eqref{eq:linear_subsys}, it is sufficient to show that (Cd.1) and (Cd.3) hold for all $x'\in\mathbb{R}^n$ with $|x'|=\beta$ by showing~\eqref{1} and~\eqref{2} hold for all $z'_i$ with $i\in[1,2^n]$.
Therefore, we are able to complete the proof by combining~\eqref{1},~\eqref{1.5}, and~\eqref{2}.
$\hfill\blacksquare$

\subsection{Proof of Theorem~\ref{thm:c-basedapp},~\ref{thn:cbasedcloss},~\ref{thm:e-basedapp}, and~\ref{thn:ebasedcloss}}\label{proof:section4.2}
We first show the results for Theorem~\ref{thm:c-basedapp}.
 
{\bf Proof of Theorem~\ref{thm:c-basedapp}}
First, we show the existence of $i\in\mathbb{N}$ such that~\eqref{stop_ro} holds.
Accordingly, we discuss two cases: 
\begin{enumerate}
	\item In case that $\underline{I}_i=\emptyset$ for some $i\in\mathbb{N}$, then $\forall i'\geq i$, one gets $\underline{I}_{i'}=\emptyset$, since $(\emptyset)_{\gamma}=\emptyset$  for any $\gamma\in\mathbb{R}_{>0}$ such that~\eqref{stop_ro} holds.
	\item In case that $\underline{I}_i\neq\emptyset$ for all $i\in\mathbb{N}$, one can verify from Theorem~\ref{thm:convergence} that for any $\gamma\in\mathbb{R}_{>0}$, there exists $i\in\mathbb{N}$ such that for all $i'\geq i$,  $\mathsf{d}_H(\underline{I}^*,\underline{I}_{i'})<\gamma$.
	Additionally, considering the computation of $\underline{I}_i$, $i\in\mathbb{N}$ as in~\eqref{iter_p}, one can verify that $\underline{I}_0\supseteq \underline{I}_1 \supseteq \ldots \supseteq \underline{I}_i \supseteq \ldots$.
	Therefore, we have $\underline{I}_{i}\subseteq(\underline{I}_{i'})_{\gamma}$.
\end{enumerate}
Thus, we conclude the proof of the existence of $i$ by combining both cases above.
Next, we proceed with showing that $\underline{I}(\varepsilon_x,\varepsilon_u)$ as in~\eqref{eq:Icbased} is an HCI set for $\mathcal{S}\otimes\mathcal{A}$.
Here, we only discuss the case in which $\underline{I}(\varepsilon_x,\varepsilon_u)\neq\emptyset$ since $\emptyset$ is a trivial solution of an HCI set for $\mathcal{S}\otimes\mathcal{A}$.
Consider any $\underline{x}=(q,q',x)\in \underline{I}(\varepsilon_x,\varepsilon_u)$. 
Then, by definition of $\underline{I}(\varepsilon_x,\varepsilon_u)$ as in~\eqref{eq:Icbased}, there exists an $i'\in[1,n']$ such that $\underline{x}\in\underline{I}_{i_*+i'}\oplus \mathcal{N}_{i'}(\varepsilon_x,\varepsilon_u)$.
Without loss of generality, we assume that $x=x_1+x_2$, with $x_1=\underline{I}_{i^*+i'}(q,q')$ and $x_2 \in  \mathcal{N}_{i'}(\varepsilon_x,\varepsilon_u)$.
On one hand, there exists $u_2\in \varepsilon_u\mathbb{B}^m$ such that $x'_2 := Ax_2+Bu_2\in \mathcal{N}_{i'-1}(\varepsilon_x,\varepsilon_u)$.
On the other hand, let $\underline{x}_1=(q,q',x_1)$.
Considering the iteration in~\eqref{iter_po}, there exists $u_1\in U - \varepsilon_u\mathbb{B}^m$ such that for all $w\in W$, $\underline{x}'_1:= (q',q'',x'_1)\in\underline{I}_{i_*+i'-1}(q',q'')$ hold, with $x'_1=Ax_1+Bu_1+w$ and $(q',L(x'_1),q'')\in\delta$.
Then, one can readily verify that for all $w\in W$, there exists $u = u_1+u_2\in U$ such that $\underline{x}'\in\underline{I}_{i_*+i'-1}\oplus \mathcal{N}_{i'-1}(\varepsilon_x,\varepsilon_u)$ for all $\underline{x}'=\underline{f}(\underline{x},u,w)$.
Now, we have the following two cases regarding different $i'$:
\begin{enumerate}
	\item (Case 1) If $i'\geq 2$, one has $\underline{x}'\in\underline{I}(\varepsilon_x,\varepsilon_u)$ by definition of $\underline{I}(\varepsilon_x,\varepsilon_u)$; 
	\item (Case 2) If $i'=1$, then according to~\eqref{stop_ro}, one gets $\underline{x}'\in \underline{I}_{i_*}\subseteq (\underline{I}_{i_*+n})_{\gamma}$.
	Additionally, considering~\eqref{eq:Hms} and Lemma~\ref{lem:cxcu}, $\gamma\mathbb{B}^n\subseteq \mathcal{N}_n(\varepsilon_x,\varepsilon_u)$ implies that $(\underline{I}_{i^*+n})_{\gamma}\subseteq \underline{I}_{i^*+n} \oplus\mathcal{N}_n(\varepsilon_x,\varepsilon_u)$.
	Therefore, $\underline{x}'\in\underline{I}(\varepsilon_x,\varepsilon_u)$ holds.
\end{enumerate}
Combining Case 1 and Case 2, one can verify that $\underline{I}(\varepsilon_x,\varepsilon_u)$ is an HCI set for $\mathcal{S}\otimes\mathcal{A}$ according to Definition~\ref{def:HCI}.$\hfill\blacksquare$

{\bf Proof of Theorem~\ref{thn:cbasedcloss}}
Consider any $\rho\in\mathbb{R}_{>0}$.
Here, we assume that $\underline{I}^*_\rho\neq\emptyset$, since~\eqref{eq:cbasedgood} holds trivially when $\underline{I}^*_\rho=\emptyset$.
For the following discussion, we define
\begin{equation*}
(S\otimes\mathcal{A})_{(-\varepsilon_x,-\varepsilon_u)}\!:=\! (\underline{X}_{-\varepsilon_x}, (\underline{X}_0)_{-\varepsilon_x},\underline{U}-\varepsilon_u\mathbb{B}^m,\underline{W},\underline{f}).
\end{equation*}
Then, we show that the assertion of Theorem~\ref{thn:cbasedcloss} holds if $\gamma = \text{min}(\rho/c_x,\rho/c_u)$.
If $\gamma = \rho/c_x$, this implies that $c_u\leq c_x$.
Consider the maximal HCI set $\underline{I}^*_{(\varepsilon_x,\varepsilon_u)}$ for the product system $(S\otimes\mathcal{A})_{(-\varepsilon_x,-\varepsilon_u)}$.
On one hand, one has 
\begin{equation}
\underline{I}^*_\rho\subseteq\underline{I}^*_{(\varepsilon_x,\varepsilon_u)},\label{eq:help1}
\end{equation}
according to the definition of  $(S\otimes\mathcal{A})_{-\rho}$, since $\varepsilon_x=\rho$ and $\varepsilon_u\leq\rho$.
On the other hand, in the view of the definition of an HCI set and the iteration as in~\eqref{iter_po}, one has 
\begin{equation}
\underline{I}^*_{(\varepsilon_x,\varepsilon_u)}\subseteq \underline{I}_{i},\label{eq:help2}
\end{equation}
for all $i\in\mathbb{N}$, with $\underline{I}_i$ being obtained through the iteration as in~\eqref{iter_po}.
Then, one can readily see that~\eqref{eq:cbasedgood} holds according to the definition of $\underline{I}(\varepsilon_x,\varepsilon_u)$ as in~\eqref{eq:Icbased}.

If $\gamma = \rho/c_u$, we can similarly show that~\eqref{eq:cbasedgood} holds.
The key insight is that $\gamma = \rho/c_u$ implies $c_x\leq c_u$, and therefore one has $\varepsilon_x\leq\rho$ and $\varepsilon_u=\rho$ for $(S\otimes\mathcal{A})_{(-\varepsilon_x,-\varepsilon_u)}$.
Then, we also have~\eqref{eq:help1} and~\eqref{eq:help2}, which completes the proof.
$\hfill\blacksquare$

{\bf Proof of Theorem~\ref{thm:e-basedapp}}
The existence of $i\in\mathbb{N}$ such that~\eqref{stop_ri} holds can be proved similarly to the existence of $i$ in Theorem~\ref{thm:c-basedapp}.
Therefore, we proceed with showing that $\underline{I}(\varepsilon)$ in~\eqref{eq:HCIebased} is an HCI set for $\mathcal{S}\otimes\mathcal{A}$.
Here, we only discuss the case in which $\underline{I}(\varepsilon)\neq\emptyset$ since $\emptyset$ is a trivial solution of an HCI set for $\mathcal{S}\otimes\mathcal{A}$.
On one hand,~\eqref{stop_ri} implies that $\forall q,q'\in Q$ with $\underline{I}_{i^*}(q,q')\neq\emptyset$, $\underline{I}_{i^*}(q,q') \subseteq \underline{I}_{i^*+1}(q,q') \oplus \varepsilon\mathbb{B}^n$ hold.
Hence, one has $(\underline{I}_{i^*})_{-\varepsilon}\subseteq \underline{I}_{i^*+1}$.
On the other hand,~\eqref{iter_pi} shows that $\forall \underline{x}:=(q,q',x')\in\underline{I}_{i^*+1}$, $\forall w \in W+\varepsilon\mathbb{B}^n$, $\exists u \in U$ such that $\underline{x}'\in\underline{I}_{i^*}$ holds, with $\underline{x}'=\underline{f}(\underline{x},u,w)$.
This indicates that $\forall \underline{x}:=(q,q',x')\in\underline{I}_{i^*+1}$ and $\forall w' \in W$, $\exists u \in U$ such that we have $\underline{x}''\in(\underline{I}_{i^*})_{-\varepsilon}\subseteq \underline{I}_{i^*+1}$,
with $\underline{x}''=\underline{f}(\underline{x},u,w')$.
Therefore, $\underline{I}_{i^*+1}$ is an HCI set for $\mathcal{S}\otimes\mathcal{A}$ according to Definition~\ref{def:HCI}, which completes the proof. $\hfill\blacksquare$

{\bf Proof of Theorem~\ref{thn:ebasedcloss}}
Consider any $\rho\in\mathbb{R}_{>0}$.
If $\underline{I}^*_\rho=\emptyset$,~\eqref{eq:ebasedgood} holds trivially.
Therefore, we focus on the case in which $\underline{I}^*_\rho\neq\emptyset$.
In the rest of this proof, we show that the assertion of Theorem~\ref{thn:ebasedcloss} holds with 
\begin{equation}
\varepsilon = \text{min}(\frac{\rho}{n'c_x},\frac{\rho}{n'c_u}), \label{proof:vareps}
\end{equation}
in which $c_x$, $c_u$, and $n'$ are those in Corollary~\ref{col:cxcu} such that~\eqref{eq:cd1}-\eqref{eq:cd3} hold.
To this end, we define a set
\begin{align}
\underline{X}':= \bigcup_{i'\in [1,n']} (\underline{I}^*_\rho\oplus  \mathcal{N}_{i'}(\varepsilon_x,\varepsilon_u)),\label{eq:helpset}
\end{align}
in which $\varepsilon_x$ and $\varepsilon_u$ are computed based on $\gamma = \varepsilon$ as in Lemma~\ref{lem:cxcu}, with $\varepsilon$ as in~\eqref{proof:vareps}.
Accordingly, one can verify 
\begin{align}
\varepsilon\mathbb{B}^n\subseteq\mathcal{N}_{n'}(\varepsilon_x,\varepsilon_u),\label{eq:001}
\end{align}
by leveraging Lemma~\ref{lem:cxcu}.
Moreover,one gets $\mathcal{N}_{i'}(\varepsilon_x,\varepsilon_u)\subseteq\varepsilon_x\mathbb{B}^n$ according to~\eqref{eq:seq}, $\varepsilon_x\mathbb{B}^n\subseteq\frac{\rho}{n'}\mathbb{B}^n$ for all $i'\in[1,n']$ according to~\eqref{proof:vareps}, and $\underline{I}^*_\rho\subseteq(\underline{I}_0)_{-\rho}$ according to the definition of HCI sets as in Definition~\ref{def:HCI}.
Therefore, one has
\begin{equation}
\underline{X}'\subseteq (\underline{I}_0)_{-\rho}\oplus (n'\times\frac{\rho}{n'}\mathbb{B}^n)=\underline{I}_0,\label{proofI0}
\end{equation}
with $\underline{I}_0$ as in~\eqref{iter_pi}.
Now, we start proving Theorem~\ref{thn:ebasedcloss}.

Consider any $\underline{x}:=(q,q',x)\in\underline{X}'$.
Without loss of generality, we assume that $x = \tilde{x}+\sum_{i=1}^{n'}x_i$, with $\tilde{x}\in\underline{I}^*_{\rho}(q,q')$, and $x_i\in \mathcal{N}_{i}(\varepsilon_x,\varepsilon_u)$ for all $i\in[1,n']$.
Since $\tilde{\underline{x}}:=(q,q',\tilde{x})\in\underline{I}^*_{\rho}$, then $\exists \tilde{u}\in \underline{U}-\rho\mathbb{B}^m$ such that for all $w\in\underline{W}$, we get $(q',q'',\tilde{\underline{x}}'):=\underline{f}(\tilde{\underline{x}},u,w)\in\underline{I}^*_{\rho}$, with $\tilde{\underline{x}}' = A\tilde{\underline{x}}+B\tilde{u}+w$ and $(q',L(\tilde{\underline{x}}'),q'')\in\delta$.
Accordingly, considering~\eqref{eq:001}, there also exists $\tilde{u}\in \underline{U}-\rho\mathbb{B}^m$ such that for all $w'\in\underline{W}+\varepsilon\mathbb{B}^n$, 
\begin{align}
(q',\tilde{q}'',\tilde{\underline{x}}''):=\underline{f}(\tilde{\underline{x}},u,w')\in\underline{I}^*_{\rho}\oplus  \mathcal{N}_{n'}(\varepsilon_x,\varepsilon_u),\label{eq:help21}
\end{align} 
hold, with $\tilde{\underline{x}}'' = A\tilde{\underline{x}}+B\tilde{u}+w'$ and $(q',L(\tilde{\underline{x}}''),\tilde{q}'')\in\delta$.
Moreover, according to Definition~\ref{def:seq:Ni}, for any $x_i\in \mathcal{N}_{i}(\varepsilon_x,\varepsilon_u)$ with $i\in[1,n']$, there exists $u_i\in\varepsilon_u\mathbb{B}^m$ such that
\begin{equation}
Ax_i+Bu_i\in\mathcal{N}_{i-1}(\varepsilon_x,\varepsilon_u).\label{eq:help22}
\end{equation}
Combining~\eqref{eq:help21} and~\eqref{eq:help22}, one can readily see that for any $\underline{x}:=(q,q',x)\in\underline{X}'$, for all $w'\in\underline{W}+\varepsilon\mathbb{B}^n$, we get $\underline{x}':=(q',q'',x')\in\underline{X}'$, with $x'= Ax+Bu+w'$, $(q',L(x'),q'')\in\delta$, and $u=\tilde{u}+\sum_{i=1}^{n'}u_i$.
Additionally, since $\gamma\leq\frac{\rho}{n'c_u}$ according to~\eqref{proof:vareps}, we obtain $\varepsilon_u\leq\frac{\rho}{n'}$ and as a result $u\in\underline{U}$.
Hence, considering~\eqref{proofI0}, one can readily conclude that the set $\underline{X}'$ is an HCI set for a product $S'\otimes\mathcal{A}$ as defined in Definition~\ref{syn_prod}, with $S'=(X,X_0,U,W+\varepsilon\mathbb{B}^n,f)$, and hence, one gets
\begin{equation}
\underline{X}'\subseteq \underline{I}^*(\varepsilon),\label{help1}
\end{equation}
with $\underline{I}^*(\varepsilon)$ being the maximal HCI set of $S'\otimes\mathcal{A}$.
Moreover, according to~\eqref{eq:helpset}, one can readily see that $\underline{I}^*_\rho\subseteq\underline{X}'\subseteq \underline{I}^*(\varepsilon)$, which completes the proof, since $\underline{I}^*(\varepsilon) \subseteq \underline{I}(\varepsilon)$ considering~\eqref{iter_p},~\eqref{stop_r}, and~\eqref{iter_pi}.$\hfill\blacksquare$

\subsection{Proof of Theorem~\ref{thm:upperbound} and Corollary~\ref{thm:complexity}}\label{proof:section5}
To prove Theorem~\ref{thm:upperbound}, the following proposition is required.
\begin{proposition}\label{prop:help}
	Given P-collections $\mathcal{U}_1$ and $\mathcal{U}_2$, one has
	\begin{align}
	\mathsf{larg}(\mathcal{U}_1\cap \mathcal{U}_2)&\leq \mathsf{larg}(\mathcal{U}_1)+\mathsf{larg}(\mathcal{U}_2),\label{eq:prop1}\\
	\mathsf{num}(\mathcal{U}_1\cap \mathcal{U}_2)&\leq  \mathsf{num}(\mathcal{U}_1)\mathsf{num}( \mathcal{U}_2),\label{eq:prop2}\\
	\mathsf{larg}(pre(\mathcal{U}_1)) &\leq \tilde{g}_S(p),\label{eq:prop3}\\
	\mathsf{num}(pre(\mathcal{U}_1))&\leq \mathsf{num}(\mathcal{U}_1),\label{eq:prop4}
	\end{align}\vspace{-0.1cm}
	\noindent
	in which $\mathsf{larg}(\cdot)$ and $\mathsf{num}(\cdot)$ are defined in~\eqref{eq:hyperplane_num} and~\eqref{eq:poly_num}, respectively;
	$pre(\cdot)$ is as in~\eqref{eq:pre_orig}, with exogenous disturbance set $W =\{\mathbf{0}_n\}$;
	$\tilde{g}_S(\cdot)$ is as in~\eqref{gS}, and $p= \max_{a\in[1,\mathsf{N}_c]} \mathsf{numh}(\mathcal{P}_a)$, with $\mathcal{U}_1 =\cup_{a=1}^{\mathsf{N_c}}(\mathcal{P}_a)$.
\end{proposition}
\vspace{0.1cm}
{\bf Proof of Proposition~\ref{prop:help}} 
~\eqref{eq:prop1} and~\eqref{eq:prop2} hold trivially according to how the intersection between two P-collection is computed, and~\eqref{eq:prop3} holds according to the definition for $\tilde{g}_S(\cdot)$.
As for~\eqref{eq:prop4}, one can verify that
\begin{small}
	\begin{align*}
	\mathsf{num}(pre(\mathcal{U}_1))\!=\!\mathsf{num}(\cup_{a=1}^{\mathsf{N}_c}pre(\mathcal{P}_a))\!\leq \!\cup_{a=1}^{\mathsf{N}_c}\mathsf{num}(pre(\mathcal{P}_a))\!\leq\! \mathsf{N_c}.
	\end{align*}
\end{small}\noindent
Note that the last inequality holds since $pre(\mathcal{P}_a)$ is still a polytope given $\mathcal{P}_a$ is polytope~\cite[Section 3.3.3]{Kerrigan2001Robust}.
$\hfill\blacksquare$

{\bf Proof of Theorem~\ref{thm:upperbound}}  
Here, we show~\eqref{eq:proff_compl1} and~\eqref{eq:proff_compl2} by induction.
When $i=1$, for any $q,q',q''\in Q_{rd}$ for which $\exists \sigma_1, \sigma_2\in\Pi$ s.t. $(q,\sigma_1,q')\in \delta$ and $(q',\sigma_2,q'')\in \delta$, one has
	\begin{align*}
	\mathsf{num}(\underline{I}_1(q,q'))&=\!\!\!\!\!\! \sum_{q''\in Q_{rd}}\!\!\! \mathsf{num}\Big(\underline{I}_0(q,q')\cap pre(\underline{I}_0(q',q''))\Big)\\
	&\leq\!\!\!\!\!\! \sum_{q''\in Q_{rd}}\!\!\! \mathsf{num}(\underline{I}_0(q,q'))\mathsf{num}(pre(\underline{I}_0(q',q''))\tag{c1}\\
	&\leq \sum_{q''\in Q_{rd}} \mathsf{M}^2\leq \alpha \mathsf{M}^2;\tag{c2}\\
	\mathsf{larg}(\underline{I}_1(q,q')) &= \mathsf{larg}\Big(\underline{I}_0(q,q')\cap pre(\underline{I}_0(q',q''))\Big)\\
	&\leq \mathsf{larg}(\underline{I}_0(q,q'))+\mathsf{larg}(pre(\underline{I}_0(q',q''))\tag{c3}\\
	& \leq p' + \tilde{g}_{S}(p') \leq g^1(p')\tag{c4}.
	\end{align*}
Hence,~\eqref{eq:proff_compl1} and~\eqref{eq:proff_compl2} hold for $i=1$.
Note that (c1)-(c4) hold according to Proposition~\ref{prop:help}.
Suppose that~\eqref{eq:proff_compl1} and~\eqref{eq:proff_compl2} hold for $i=k$.
Then, for $i=k+1$, one has
	\begin{align*}
	\mathsf{larg}(\underline{I}_{i+1}(q,q')) &= \mathsf{larg}\Big(\underline{I}_0(q,q')\cap pre(\underline{I}_i(q',q''))\Big)\\
	&\leq \mathsf{larg}(\underline{I}_0(q,q'))+\mathsf{larg}(pre(\underline{I}_i(q',q''))\leq p' + \tilde{g}_{S}(g^i(p')) \leq g^{i+1}(p').
	\end{align*}
	\begin{align*}
	\mathsf{num}(\underline{I}_{i+1}(q,q'))&=  \!\!\!\sum_{q''\in Q_{rd}}\!\! \mathsf{num}\Big(\underline{I}_0(q,q')\cap pre(\underline{I}_{i}(q',q''))\Big)\\
	&\leq \!\!\!\sum_{q''\in Q_{rd}}\!\! \mathsf{num}(\underline{I}_0(q,q'))\mathsf{num}(pre(\underline{I}_{i}(q',q''))\leq  \!\!\!\sum_{q''\in Q_{rd}} \mathsf{M}\alpha^i\mathsf{M}^{i+1}\leq \alpha^{i+1} \mathsf{M}^{i+2};
	\end{align*}
Therefore,~\eqref{eq:proff_compl1} and~\eqref{eq:proff_compl2} also hold for $i=k+1$, which completes the proof.
$\hfill\blacksquare$

{\bf Proof of Corollary~\ref{thm:complexity}}  
In the following discussion, considering a P-collection $\mathcal{U}=\cup_{a=1}^{\mathsf{N}_c}\mathcal{P}_a$, we denote by $\mathsf{numh}_{c}(\mathcal{U}):=\sum_{a=1}^{\mathsf{N}_c}\mathsf{numh}(\mathcal{P}_a)$ the \emph{total number of hyperplanes defining the polytopes within $\mathcal{U}$}. 
Then, based on~\eqref{eq:proff_compl1} and~\eqref{eq:proff_compl2}, one has
\begin{align*}
\mathsf{numh}_{c}(\underline{I}_i(q,q'))\leq\mathsf{num}(\underline{I}_i(q,q'))\mathsf{larg}(\underline{I}_i(q,q'))\leq \alpha^i\mathsf{M}^{i+1}g^i(p').
\end{align*}
Therefore, $I_i$ contains at most $|\delta|\alpha^i\mathsf{M}^{i+1}g^i(p')$ hyperplanes.
	Meanwhile, the parameters of these hyperplanes can be stored in a $|\delta|\alpha^i\mathsf{M}^{i+1}g^i(p')$-by-$(n+1)$ matrix.
	Hence, ~\eqref{eq:spacecomp} is a valid upper bound for the space complexities of Algorithm~\ref{alg}.
	Next, we proceed with showing that~\eqref{eq:timecomp} is a valid upper bound for the time complexity of Algorithm~\ref{alg}.
	First, considering~\eqref{eq:proff_compl1},~\eqref{eq:mathsfM},~\eqref{eq:prop3}, and~\eqref{eq:prop4}, one has
	\begin{align*}
	\mathsf{num}(pre(\underline{I}_{i-1}(q,q')))\leq\mathsf{num}(\underline{I}_{i-1}(q,q'))\leq\alpha^{i-1}\mathsf{M}^{i},\\
	\mathsf{larg}(pre(\underline{I}_{i-1}(q,q')))\leq \tilde{g}_{S}(g^{i-1}(p')).
	\end{align*}
	Accordingly, in the worst case, one needs to compute the intersection of two P-collections, which contains $\mathsf{M}$ and $\alpha^{i-1}\mathsf{M}^{i}$ polytopes, respectively, to obtain $\underline{I}_0\cap \textbf{P}(\underline{I}_i)$.
	Therefore, the worst-case computation time for computing $\underline{I}_0\cap \textbf{P}(\underline{I}_i)$ is $|\delta|\alpha^{i-1}\mathsf{M}^{i+1}c_2\big(p',\tilde{g}_{S}(g^{i-1}(p'))\big)$ considering the definition of $c_2$ and $|\delta|$.
	Then, one can readily verify that~\eqref{eq:timecomp} is a valid upper bound for the time complexity of Algorithm~\ref{alg} by considering the definitions of $c_1$ and $c_3$. $\hfill\blacksquare$

\end{document}